\documentclass[aps,prd,amsmath,twocolumn,preprintnumbers]{revtex4}
\usepackage{graphicx}

\begin{document}

\date{\today}

\title{Mass loss and longevity of gravitationally bound oscillating
scalar lumps (oscillatons) in $D$-dimensions}

\author{Gyula Fodor$^1$, P\'eter Forg\'acs$^{1,2}$,
 M\'ark Mezei$^{3,4}$}
\affiliation{$^1$MTA RMKI, H-1525 Budapest 114, P.O.Box 49, Hungary,\\
$^2$LMPT, CNRS-UMR 6083, Universit\'e de Tours, Parc de Grandmont,
37200 Tours, France\\
$^3$Center for Theoretical Physics, Massachusetts Institute of Technology,
Cambridge, Massachusetts 02139,USA \\
$^4$Institute for Theoretical Physics, E\"otv\"os University,
 H-1117 Budapest, P\'azm\'any P\'eter s\'et\'any 1/A, Hungary,\\
}

\begin{abstract}
  Spherically symmetric {\sl oscillatons} (also referred to as oscillating
  soliton stars) i.e.\ gravitationally bound oscillating scalar lumps
   are considered in theories
  containing a massive self-interacting real scalar field coupled to
  Einstein's gravity in $1+D$ dimensional spacetimes.  Oscillations
  are known to decay by emitting scalar radiation with a characteristic
  time scale which is, however, extremely long, it can be comparable even to
  the lifetime of our universe.  In the limit when the central density
  (or amplitude) of the oscillaton tends to zero (small-amplitude
  limit) a method is introduced to compute the transcendentally small
  amplitude of the outgoing waves.  The results are illustrated in
  detail on the simplest case, a single massive free scalar field
  coupled to gravity.

\end{abstract}

\preprint{MIT-CTP 4104}

\maketitle

\section{Introduction}

Numerical simulations of Seidel and Suen\cite{Seidel1} have revealed
that spatially localized, extremely long living, oscillating
configurations evolve from quite general initial data in the
spherically symmetric sector of Einstein's gravity coupled to a a
free, massive real Klein-Gordon field.  For example, they observed
that initially Gaussian pulses evolve quickly into configurations
which appear to be time-periodic.  It has been already noted in
Ref.\cite{Seidel1}, that the resulting objects may not be strictly
time-periodic, rather they may evolve on a secular time scale many
orders of magnitude longer than the observed oscillation period.
These interesting objects were first baptized "oscillating soliton
stars" in Ref. \cite{Seidel1}, but somewhat later the same objects
have been referred to as "{\sl oscillatons}" by the same authors
\cite{Seidel2}.  This latter name has been by now widely adopted, and
we shall also stick to its usage throughout this paper.  It has been
observed in the numerical simulations of Ref.\ \cite{Seidel1} that
oscillatons are {\sl stable} during the time evolution.  Moreover it
has been argued in Ref.\cite{Seidel2} that oscillatons do form in
physical processes through a dissipationless gravitational cooling
mechanism, making them of great physical importance.  For example
oscillatons would be good candidates for dark matter in our Universe.

On the other hand, stimulated by the seminal work of Dashen,
Hasslacher and Neveu in the one-dimensional $\phi^4$-theory
\cite{Dashen}, numerical simulations have revealed that in an
impressive number of scalar field theories spatially localized
structures --{\sl oscillons}-- form {\sl from generic initial data}
which become very closely time periodic, and live for very long times
\cite{BogMak2,CopelGM95,Chris,PietteZakr98,Honda,Hindmarsh-Salmi06,
  SafTra,fggirs,sicilia2}.  These objects oscillate nearly
periodically in time, resembling ``true'' (i.e.\ time-periodic)
breathers.  An oscillon possesses a ``radiative'' tail outside of its
core region where its energy is leaking continuously in form of
(scalar) radiation.  Therefore a simple approximate physical picture
of a sufficiently small-amplitude oscillon is the that of a ``true''
breather whose frequency is increasing on a secular time scale since
the amplitude of the outgoing radiation is much smaller than that of
the core.  It has been shown in Refs.\ \cite{FFGR}, \cite{FFHL}, that
slowly radiating oscillons can be well described by a special class of
exactly time-periodic ``quasibreathers'' (QB). Being time periodic,
QBs are easier to describe mathematically by ordinary Fourier analysis
than the long time asymptotics of oscillons. A QB possesses a
localized core in space (just like true breathers) which approximates
that of the corresponding oscillon very well, but in addition it has a
standing wave tail whose amplitude is {\sl minimized}.  This is a
physically motivated condition, which heuristically singles out
``the'' solution approximating a true breather as well as possible,
for which this amplitude would be identically zero.  The amplitude of
the standing wave tail of a QB is closely related to that of the
oscillon radiation, therefore its computation is of prime
interest. Roughly speaking ``half'' of the standing wave tail
corresponds to incoming radiation from spatial infinity.  It is the
incoming radiation that maintains the time periodicity of the QBs by
compensating the energy loss through the outgoing waves.  In a series
of papers \cite{FFHL,FFHM,moredim} a method has been developed to
compute the leading part of the exponentially suppressed tail
amplitude of QBs, in a large class of scalar theories in various
dimensions, in the limit when the QB core amplitude is small.
Although oscillons continuously loose energy through radiation, many
of them are remarkably stable.  The longevity and the ubiquity of
oscillons make them of potentially great physical interest
\cite{Kolb:1993hw,Khlopov,Broadhead:2005hn,Gleiser:2007ts,Borsanyi2}.
Quite importantly oscillons also appear in the course of time
evolution when other fields, e.g. vector fields are present
\cite{Farhi05,Graham07a,Graham07b}.  There is little doubt that
oscillons and oscillatons are closely related objects.

The basic physical mechanism for the anti-intuitively slow radiation
of oscillons is that the lowest frequency mode of the scalar field is trapped
below the mass threshold and only the higher frequency modes
are coupled to the continuum. 

In this paper we generalize the method of Refs.\
\cite{FFHL,FFHM,moredim} to compute the mass loss of spherically
symmetric oscillations induced by scalar radiation in the limit of
small oscillaton amplitudes, $\varepsilon$, in $1+D$ dimensional
spacetimes.  These methods have been succesfully applied to
$D$-dimensional scalar field theories coupled to a dilaton field
\cite{dilaton}. Numerous similarities exist between coupling a theory
to a dilaton field and to gravitation: the field configurations are of
$\varepsilon^2$ order and the lowest order equations determining the
profiles are the Schr\"odinger-Newton equations. The stability pattern
is also analogous.  Despite these similarities between the dilaton and
the gravitational theory there are some technical and even some
conceptual differences.  Since there is no timelike Killing vector
neither for oscillatons nor for the corresponding QBs, already the
very definition of mass and mass loss is less obvious than in flat
spacetime.  Another conceptual issue is that the spacetime of a
time-periodic QB is not asymptotically flat, which is related to the
fact that the ``total mass" of a QB is infinite. In the case of
spherical symmetry considered in this paper a suitable local mass
function is the Misner-Sharp energy and the mass loss can be defined
with aid of the Kodama vector. The issue of the precise asymptotics of
spacetimes can be sidestepped in the limit $\varepsilon\to0$ by
considering only a restricted, approximatively flat spacetime region
containing the core of the QB (having a size of order ${\cal
  O}(1/\varepsilon)$) and part of its oscillating tail. We find that
to leading order in the $\varepsilon$ expansion the oscillaton core is
determined by the $D$-dimensional analogues of the
Schr\"odinger-Newton equations 
\cite{Ruffini,Friedberg,Ferrell,Moroz,Tod}
independently of the self-interaction potential. It turns out that
exponentially localized oscillatons exist for $2<D<6$.  These findings
show a striking similarity to dilaton-scalar theories as found in
Ref.\ \cite{dilaton}.  In the case of spherically symmetric
oscillatons no gravitational radiation is expected due to Birkhoff's
theorem. The mass loss of spherically symmetric oscillatons is
entirely due to scalar radiation.

The following simple formula gives the
mass loss of a small-amplitude oscillaton in $D$ spatial dimensions:
\begin{equation}\label{e:symradlaw-intr}
\frac{\mathrm{d} M}{\mathrm{d} t}=
-\frac{c_1}{m^{D-3}\varepsilon^{D-1}}
\exp\left(-\frac{c_2}{\varepsilon}\right)\,,
\end{equation}
where $m$ denotes the mass of the scalar field, $c_1$ is a
$D$-dependent constant, while $c_2$ depends on both $D$ and the
self-interaction scalar potential.  The numerical values of
$c_1\,,c_2$ in the Einstein-Klein-Gordon (EKG) theory for spatial
dimensions $D=3,4,5$ are given in Table \ref{ctable}. We also compute
and tabulate the most important physical properties of oscillatons in
the EKG theory (their mass as a function of time, their radii).  We
would like to stress, that the method is applicable for oscillatons in
scalar theories with any self-interaction potential developable into
power series.

In the seminal work of Don N.~Page \cite{Page} both the classical and
quantum decay rate of oscillatons has been considered for the case of
free massive scalars in the EKG theory (for $D=3$).  We agree with the
overall qualitative picture of the oscillaton's mass loss found in
Ref.\cite{Page}, however, there are also some differences in the
quantitative results. For example, the amplitude of the outgoing wave
(related to $\sqrt{c_1}$) found by our method differs significantly
from that of Ref.\cite{Page}.  The main source of this discrepancy is
due to the fact that this amplitude is given by an infinite series in
the $\varepsilon$ expansion, where all terms contribute by the same
order, whereas in the estimate of Ref.\cite{Page} only the lowest
order term in this series has been used.  Our methods which are based
on the work of Segur-Kruskal \cite{SK} avoid this difficulty
altogether, moreover for the class of self-interaction potentials
containing only even powers of the scalar field, $\Phi$, the radiation
amplitude can be computed analytically using Borel summation.

We now give a lightning review on previous results scattered in the
literature on oscillatons in 3+1 dimensions.  For a given scalar field
mass, $m$, there is a one-parameter family of oscillatons,
parametrized, for example, by the central amplitude of the field,
$\Phi_{\rm c}$. As $\Phi_{\rm c}$ increases from small values, the
mass of the oscillaton, $M$, is getting larger, while the radius of
the configuration decreases. For a critical value of the central
amplitude, $\Phi_{\rm crit}$, a maximal mass configuration is reached.
Oscillatons with central amplitudes $\Phi_{\rm c}>\Phi_{\rm crit}$ are
unstable \cite{Seidel1}. This behavior is both qualitatively and
quantitatively very similar to that of boson stars
\cite{Ruffini,Feinblum,Kaup}, and also to the behavior of white
dwarfs and neutron stars \cite{Harrison}. For reviews of the vast
literature on boson stars see for example, Refs.\ \cite{Jetzer} and
\cite{SchunkMielke}. In Refs.\ \cite{Hawley,Hawleyphd} a one-parameter
family of oscillaton-type solutions in an Einstein-scalar theory with
two massive, real scalar fields has been presented, which are
essentially transitional states between boson stars and oscillatons.

The interaction of weak gravity axion field oscillatons with white
dwarfs and neutron stars have been discussed in
\cite{Iwazaki1,Iwazaki2}, proposing a possible mechanism for gamma ray
bursts \cite{Iwazaki3}. Since for very low mass scalar fields
oscillatons may be extremely heavy, it has been suggested that they
may be the central object of galaxies \cite{MatosGuzman}, or form the
dark matter galactic halos
\cite{Alcubgalactic,Susperregi,Guzman1,Guzman2,Hernandez,Guzman3,Bernal}.

Qualitatively good results for various properties of oscillatons has
been obtained by Ure\~na-L\'opez \cite{Lopez}, truncating the Fourier
mode decomposition of the field equations at as low order as
$\cos(2\omega t)$, where $\omega$ is the fundamental frequency. Then
the space and time dependence of the scalar field separates as
$\Phi(t,r)=\Phi_1(r)\cos(\omega t)$. Oscillatons with nontrivial
self-interaction potentials have also been studied in \cite{Lopez},
indicating that similarly to boson stars, the maximal mass can be
significantly larger than in the Klein-Gordon case.

The Fourier mode equations have been studied in \cite{LopezMatos} up
to orders $\cos(10\omega t)$.
The obtained value of the maximal mass by this higher order
truncation is $0.607/m$ in Planck units.
For small-amplitude nearly Minkowskian configurations spatial
derivatives are also small, and in Ref.\ \cite{LopezMatos} (and
independently in \cite{Kichena2}) it has been demonstrated that
such nearly flat oscillatons can be described by a pair of coupled differential
equations, the so called time independent the Schr\"odinger-Newton
equations \cite{Ruffini,Friedberg,Ferrell,Moroz,Tod}. These equations also
describe the weak gravity limit of boson stars. For quantum mechanical
motivations leading to the Schr\"odinger-Newton equations see
\cite{Diosi,Penrose}.

The time evolution of perturbed oscillatons has been investigated in
detail by \cite{Alcub}. For each mass smaller than the maximal
oscillaton mass there are two oscillaton configurations. The one with
the larger radius is a stable S-branch oscillaton, and the other is an
unstable U-branch oscillaton. Moderately perturbed S-branch
oscillatons vibrate with a low frequency corresponding to a
quasinormal mode. Perturbed U-branch oscillatons collapse to black
holes if the perturbation increases their mass, otherwise they migrate
to an S-branch oscillaton. Actually, U-branch oscillatons turn out to
be the critical solutions for type I critical collapse of massive
scalar fields \cite{Brady}. Corresponding apparently periodic objects
also form in the critical collapse of massive vector fields
\cite{Garfinkle}.

There are also excited state oscillatons, indexed by the nodes of the
scalar field. The instability and the decay of excited state
oscillatons into black holes or S-branch oscillatons is described in
\cite{Balak}. The evolution of oscillatons on a full 3D grid has been
also performed in \cite{Balak}, calculating the emitted gravitational
radiation.  Since $f(R)$ gravity theories are equivalent to ordinary
general relativity coupled to a real scalar field, oscillatons
naturally form in these theories as well \cite{Obregon}. The geodesics
around oscillatons has been investigated in \cite{Becerril}.

The plan of the paper is the following.  In Section \ref{Sec:Scalars}
the general formalism concerning a classical real scalar field coupled
to gravitation in $D$ dimensional, spherically symmetric spacetimes is
set up. In subsection \ref{sec:conf-flat} the coupled Einstein-scalar
equations are explicited in a spatially conformally flat coordinate
system. In Section \ref{sec:smallampl} the small-amplitude expansion
is presented and is carried out in detail.  In subsection \ref{sec:SN}
it is shown that in leading order one obtains the Schr\"odinger-Newton
eqs.\ in $D$ dimensions. In subsection \ref{sec:next} the next to
leading order results are given.  Subsection \ref{sec:sing} contains
an analysis of the singularities in the complexified radial variable.
In Section \ref{sec:masses} the proper mass resp.\ the total mass of
the QB core is evaluated in subsection \ref{sec:propermass} resp.\
subsection \ref{sec:totmass}.  In subsection \ref{sec:stab} a
conjecture for a criterion of oscillaton stability is formulated. In
Section \ref{sec:radlaw} the Fourier analysis of the field equations
is related to the small-amplitude expansion, and the amplitude of the
standing wave tail of the QB is determined using Borel summation
techniques.  In subsection \ref{sec:massloss} the mass loss rate of
oscillatons in the EKG theory is computed for $D=3,4,5$ and for
various values of the mass of the scalar field.

\section{Scalar field on curved background}\label{Sec:Scalars}

\subsection{Field equations}

We consider a real scalar field $\Phi$ with a self-interaction
potential $U(\Phi)$ in a $D+1$ dimensional curved spacetime with
metric $g_{ab}$. We use Planck units with $G=c=\hbar=1$. For a free
field with mass $m$ the potential is $U(\Phi)=m^2\Phi^2/2$.  The total
Lagrangian density is
\begin{equation}
\mathcal{L}=\mathcal{L}_G+16\pi\mathcal{L}_M \,,
\end{equation}
where the Einstein Lagrangian density is $\mathcal{L}_G=\sqrt{-g}\,R$,
and the Lagrangian density belonging to the scalar field is
\begin{equation}
\mathcal{L}_M=-\sqrt{-g}\left(\frac{1}{2}\Phi_{,a}\Phi^{,a}
+U(\Phi)\right) \,.
\end{equation}
Variation of the action with respect to $\Phi$ yields the wave
equation
\begin{equation}
g^{ab}\Phi_{;ab}-U'(\Phi)=0 \,, \label{eq:wave}
\end{equation}
while variation with respect to
$g^{ab}$ yields Einstein equations
\begin{equation}
G_{ab}=8\pi T_{ab} \,, \label{eq:einst}
\end{equation}
where the stress-energy tensor is
\begin{equation}
T_{ab}=\Phi_{,a}\Phi_{,b}-g_{ab}
\left(\frac{1}{2}\Phi_{,c}\Phi^{,c}+U(\Phi)\right) . \label{eq:tab}
\end{equation}
If $D=1$ then, by definition, the Einstein tensor is traceless, and
from the trace of the Einstein equations it follows that
$U(\Phi)=0$. Hence we assume that $D>1$.

We shall assume that the self-interaction potential, $U(\Phi)$, has a
minimum $U(\Phi)=0$ at $\Phi=0$, and expand its derivative as
\begin{equation}
U'(\Phi)=\sum\limits_{k=1}^{\infty}u_k\Phi^k\,,
\end{equation}
where $u_k$ are constants.  In order to get rid of the $8\pi$ factors
in the equations we introduce a rescaled scalar field and potential by
\begin{equation}
\phi=\sqrt{8\pi}\,\Phi \ , \qquad
\bar U(\phi)=8\pi U(\Phi) \,.  \label{eq:rscpu}
\end{equation}
Then
\begin{equation}
\bar U'(\phi)=\sum\limits_{k=1}^{\infty}v_k\phi^k\,,
\end{equation}
with
\begin{equation}
v_k=\frac{u_k}{(8\pi)^{(k-1)/2}} \,.
\end{equation}
The mass of the field is $m\equiv\sqrt{u_1}=\sqrt{v_1}$. If the pair
$\phi(x^c)$ and $g_{ab}(x^c)$ solves the field equations with a
potential $\bar U(\phi)$, then $\hat\phi(x^c)=\phi(\gamma x^c)$ and
$\hat g_{ab}(x^c)=g_{ab}(\gamma x^c)$, for any positive constant
$\gamma$, is a solution with a rescaled potential $\gamma^2\bar
U(\phi)$. It is sufficient to study the problem with potentials
satisfying $m^2=u_1=v_1=1$, since the solutions corresponding to an
arbitrary potential can be obtained from the solutions with an
appropriate potential with $m=1$ by applying the transformation
\begin{equation}
\phi(x^c) \to \phi(m x^c) \ , \qquad
g_{ab}(x^c) \to g_{ab}(m x^c) \ . \label{eq:trans}
\end{equation}
To simplify the expressions, unless explicitly stated, in the
following we assume $m=1$.

\subsection{Spherically symmetric $D+1$ dimensional spacetime}

We consider a spherically symmetric $D+1$ dimensional spacetime with
coordinates $x^\mu=(t,r,\theta_1,...,\theta_{D-1})$. The metric can be
chosen diagonal with components
\begin{equation}\label{eq:metrgen}
\begin{split}
g_{tt}&=-A \ , \qquad\, g_{rr}=B \ ,\\
g_{\theta_1\theta_1}&=C \ , \qquad
g_{\theta_n\theta_n}=C\prod_{k=1}^{n-1}\sin^2\theta_k \ ,
\end{split}
\end{equation}
where $A$, $B$ and $C$ are functions of temporal coordinate $t$ and
radial coordinate $r$. The nonvanishing components of the Einstein
tensor and the form of the wave equation are given in Appendix
\ref{app:einstein}.

A natural radius function, $\hat r$, can be defined in terms of the
area of the symmetry spheres in general spherically symmetric
spacetimes. In the metric \eqref{eq:metrgen} it is simply
\begin{equation}
\hat r=\sqrt{C} \,.  \label{eq:radfunc}
\end{equation}
The Kodama vector \cite{Kodama,Hayward} is defined then by
\begin{equation}
K^a=\epsilon^{ab}\hat r_{,b} \,,
\end{equation}
where $\epsilon_{ab}$ is the volume form in the $(t,r)$ plane. Choosing
the orientation such that $\epsilon_{rt}=\sqrt{AB}$ makes $K^a$ future
pointing, with nonvanishing components
\begin{equation}
K^t=\frac{\hat r_{,r}}{\sqrt{AB}} \  , \qquad
K^r=-\frac{\hat r_{,t}}{\sqrt{AB}} \,.
\end{equation}
It can be checked that, in general, the Kodama vector is divergence
free, $K^a_{\ ;a}=0$. Since contracting with the Einstein tensor,
$G^{ab}K_{a;b}=0$, the current
\begin{equation}
J_a=T_{ab}K^b
\end{equation}
is also divergence free, $J^a_{\ ;a}=0$, it defines a conserved
charge. Integrating on a constant $t$ hypersurface with a future
oriented unit normal vector $n^a$, the conserved charge is
\begin{align}
E&=\frac{2\pi^{\frac{D}{2}}}{\Gamma\left(\frac{D}{2}\right)}
\int_0^r\hat r^{D-1}\sqrt{B}\,n^aJ_a dr  \label{eq:er}\\
&=\frac{2\pi^{\frac{D}{2}}}{\Gamma\left(\frac{D}{2}\right)}
\int_0^r\frac{\hat r^{D-1}}{A}\left(T_{tt}\hat r_{,r}
-T_{tr}\hat r_{,t}\right) dr \,. \notag
\end{align}
It is possible to show \cite{Kodama,Hayward}, that $E$ agrees with the
Misner-Sharp energy (or local mass) function $\hat m$
\cite{MisnerSharp}, which can be defined for arbitrary dimensions by
\begin{equation}
\hat m=\frac{(D-1)\pi^{\frac{D}{2}}}
{8\pi\Gamma\left(\frac{D}{2}\right)}
\hat r^{D-2}\left(1-g^{ab}\hat r_{,a}\hat r_{,b}\right)
\,. \label{eq:massfunc}
\end{equation}
It can be checked by a lengthy calculation, that the derivative of the
mass function is
\begin{equation}
\hat m_{,a}=-\frac{2\pi^{\frac{D}{2}}\hat r^{D-1}}
{\Gamma\left(\frac{D}{2}\right)}
\epsilon_{ab}J^b \,.
\end{equation}
For the radial derivative follows that
\begin{equation}
\hat m_{,r}=\frac{2\pi^{\frac{D}{2}}\hat r^{D-1}}
{\Gamma\left(\frac{D}{2}\right)A}
\left(T_{tt}\hat r_{,r}
-T_{tr}\hat r_{,t}\right) \,,
\end{equation}
which, comparing with \eqref{eq:er}, gives $E=\hat m$. Since for large
$r$ the function $\hat m$ tends to the total mass, this relation will
be important when calculating the mass loss rate caused by the scalar
radiation in Section \ref{sec:massloss}.  The time derivative of the
mass function is
\begin{equation}
\hat m_{,t}=\frac{2\pi^{\frac{D}{2}}\hat r^{D-1}}
{\Gamma\left(\frac{D}{2}\right)B}
\left(T_{rt}\hat r_{,r}
-T_{rr}\hat r_{,t}\right) \,.
\end{equation}
This equation is according to the expectation, that, because of the
spherical symmetry, the mass loss is caused only by the outward energy
current of the massive scalar field. If at large distances the metric
becomes asymptotically Minkowskian, $A=B=1$, $C=r^2$ and $\hat r=r$,
then using \eqref{eq:tab} and \eqref{eq:rscpu},
\begin{equation}
\hat m_{,t}=\frac{2\pi^{\frac{D}{2}}r^{D-1}}
{\Gamma\left(\frac{D}{2}\right)}
\Phi_{,t}\Phi_{,r}=\frac{2\pi^{\frac{D}{2}}r^{D-1}}
{8\pi\Gamma\left(\frac{D}{2}\right)}
\phi_{,t}\phi_{,r} \,.  \label{eq:minkmt}
\end{equation}

\subsection{Spatially conformally flat coordinate system}\label{sec:conf-flat}

The diffeomorphism freedom of the general spherically symmetric
time-dependent metric form \eqref{eq:metrgen} can be fixed in various
ways. The most obvious choice is the use of Schwarzschild area
coordinates by setting $C=r^2$. However, as it was pointed out by Don
N.~Page in \cite{Page}, for the oscillaton problem it is more
instructive to use the spatially conformally flat coordinate system
defined by
\begin{equation}
C=r^2B \,,  \label{eq:conffl}
\end{equation}
even if some expressions are becoming longer by this choice.  As we
will see in Sec.\ \ref{sec:smallampl} and in Appendix \ref{app:schw},
inside the oscillaton the spheres described by constant Schwarzschild
$r$ coordinates are oscillating with much larger amplitude than the
constant $r$ spheres in the conformally flat coordinate system.  In
both coordinates, when the functions $A$ and $B$ tend to $1$, the
spacetime approaches the flat Minkowskian metric.

\begin{widetext}
In the spatially conformally flat coordinate system the Einstein
equations take the form
\begin{align}
(D-1)\left[\frac{D}{4B^2}\left(B_{,t}\right)^2
-\frac{A}{r^{D-1}B^{\frac{D+2}{4}}}\left(\frac{r^{D-1}B_{,r}}
{B^{\frac{6-D}{4}}}\right)_{,r}
\right]&=\left(\phi_{,t}\right)^2
+\frac{A}{B}\left(\phi_{,r}\right)^2+2A\bar U(\phi) , \label{eq:eieq1}\\
(D-1)\left[
\frac{(D-2)\left(r^2B\right)_{,r}}{4r^4A^{\frac{2}{D-2}}B^2}
\left(r^2A^{\frac{2}{D-2}}B\right)_{,r}
-\frac{1}{A^{\frac{1}{2}}B^{\frac{D}{4}-1}}
\left(\frac{B^{\frac{D}{4}-1}B_{,t}}{A^{\frac{1}{2}}}
\right)_{,t}
-\frac{D-2}{r^2}
\right]&=\left(\phi_{,r}\right)^2
+\frac{B}{A}\left(\phi_{,t}\right)^2-2B\bar U(\phi) , \label{eq:eieq2}\\
-\frac{D-1}{2}\,A^{\frac{1}{2}}\left(\frac{B_{,t}}
{A^{\frac{1}{2}}B}\right)_{,r}&=\phi_{,t}\,\phi_{,r} , \label{eq:eieq3}\\
\frac{rB}{A^{\frac{1}{2}}}\left(\frac{A_{,r}}
{rA^{\frac{1}{2}}B}\right)_{,r}
+(D-2){rB^{\frac{1}{2}}}\left(\frac{B_{,r}}{rB^{\frac{3}{2}}}
\right)_{,r}&=2
\left(\phi_{,r}\right)^2 . \label{eq:eieq4}
\end{align}
The right hand sides are equal to $2G_{tt}$, $2G_{rr}$, $G_{tr}$ and
$2(G_{\theta_1\theta_1}/r^2-G_{rr})$, respectively. The wave equation is
then
\begin{equation}
\frac{\phi_{,rr}}{B}-\frac{\phi_{,tt}}{A}
+\frac{\phi_{,r}}{2r^{2D-2}AB^{D-1}}\left(r^{2D-2}AB^{D-2}\right)_{,r}
-\frac{\phi_{,t}}{2B^{D}}\left(\frac{B^{D}}{A}\right)_{,t}
-\bar U'(\phi)=0 \,. \label{eq:wave3}
\end{equation}
\end{widetext}

\section{Small-amplitude expansion} \label{sec:smallampl}

The small-amplitude expansion procedure has been applied successfully
to describe the core region of one-dimensional flat background
oscillons in $\phi^4$ scalar theory
\cite{Dashen,SK,Kichenassamy}. Later it has been generalized for $D+1$
dimensional spherically symmetric systems in \cite{FFHL}, and to a
scalar-dilaton system in \cite{dilaton}. In this section we generalize
the method for the case when the scalar field is coupled to gravity.

\subsection{Choice of coordinates}

We are looking for spatially localized bounded solutions of the field
equations \eqref{eq:einst} for which $\phi$ is small and the metric is
close to flat Minkowskian. We use the spatially conformally flat
coordinate system defined by \eqref{eq:conffl}. It turns out, that
under this approximation, all configurations that remain bounded as
time passes are necessarily periodically oscillating in time. We
expect that similarly to flat background oscillons, the smaller the
amplitude of an oscillaton is, the larger its spatial extent
becomes. Numerical simulation of oscillatons clearly support this
expectation. Therefore, we introduce a new radial coordinate $\rho$ by
\begin{equation}
\rho=\varepsilon r \,,
\end{equation}
where $\varepsilon$ denotes the small-amplitude parameter.  We expand
$\phi$ and the metric functions in powers of $\varepsilon$ as
\begin{align}
\phi&=\sum_{k=1}^\infty\epsilon^{2k}\phi_{2k} \,,\label{eq:phiexp}\\
A&=1+\sum_{k=1}^\infty\epsilon^{2k}A_{2k} \,, \\
B&=1+\sum_{k=1}^\infty\epsilon^{2k}B_{2k} \,. \label{eq:bexp}
\end{align}
Since we intend to use asymptotically Minkowskian coordinates, where
far from the oscillaton $t$ measures the proper time and $r$ the
radial distances, we look for functions $\phi_{2k}$, $A_{2k}$ and
$B_{2k}$ that tend to zero when $\rho\to\infty$.  One could initially
include odd powers of $\varepsilon$ into the expansions
\eqref{eq:phiexp}-\eqref{eq:bexp}, however, it can be shown by the
method presented below, that the coefficients of those terms
necessarily vanish when we are looking for configurations that remain
bounded in time.

The frequency of the oscillaton also depends on its
amplitude. Similarly to the flat background case we expect that the
smaller the amplitude is, the closer the frequency becomes to the
threshold $m=1$. Numerical simulations also show this. Hence we
introduce a rescaled time coordinate $\tau$ by
\begin{equation}
\tau=\omega t \,.
\end{equation}
and expand the square of the $\varepsilon$ dependent factor $\omega$
as
\begin{equation}
\omega^2=1+\sum_{k=1}^\infty\varepsilon^{2k}\omega_{2k} \,.
\end{equation}
It is possible to allow odd powers of $\varepsilon$ into the expansion
of $\omega^2$, but the coefficients of those terms turn out to be zero
when solving the equations arising from the small-amplitude expansion.
There is a considerable freedom in choosing different parametrizations
of the small-amplitude states, changing the actual form of the
function $\omega$. The physical parameter is not $\varepsilon$ but the
frequency of the periodic states that will be given by
$\omega$. Similarly to the dilaton model in \cite{dilaton}, we will
show, that for spatial dimensions $2<D<6$ the parametrization of the
small-amplitude states can be fixed by setting
$\omega=\sqrt{1-\varepsilon^2}$.

\subsection{Leading order results}

The field equations we solve are the Einstein equations
\eqref{eq:eieq1}-\eqref{eq:eieq4}, together with the wave equation
\eqref{eq:wave3}, using the spatially conformally flat coordinate
system $C=r^2B$. The results of the corresponding calculations in
Schwarzschild area coordinates $C=r^2$ are presented in Appendix
\ref{app:schw}. Since we look for spatially slowly varying
configurations with an $\varepsilon$ dependent frequency, we apply the
$\varepsilon$ expansion in $\tau$ and $\rho$ coordinates. This can be
achieved by replacing the time and space derivatives as
\begin{equation}
\frac{\partial}{\partial t} \to \omega\frac{\partial}{\partial\tau}
\ , \qquad
\frac{\partial}{\partial r} \to
\varepsilon\frac{\partial}{\partial\rho} \,,
\end{equation}
and substituting $r=\rho/\varepsilon$.

From the $\varepsilon^2$ components of the field equations follows
that
\begin{equation}
\phi_2=p_2\cos(\tau+\delta) \,, \quad B_2=b_2 \,,
\end{equation}
where three new functions, $p_2$, $\delta$ and $b_2$ are introduced,
depending only on $\rho$. From the $\varepsilon^4$ part of
\eqref{eq:eieq3} it follows that $\delta$ is a constant. Then by a
shift in the time coordinate we set
\begin{equation}
\delta=0 \,.
\end{equation}
This shows that the scalar field oscillates simultaneously, with the
same phase at all radii.

The $\varepsilon^4$ component of the field equations yield that
\begin{align}
A_2&=a_2 \,, \\
\phi_4&=p_4\cos\tau+\frac{v_2p_2^2}{6}\left[\cos(2\tau)-3\right]
\,, \label{eq:phi4eps}\\
B_4&=b_4-\frac{p_2^2}{4(D-1)}\cos(2\tau) \,,
\end{align}
where $a_2$, $p_4$ and $b_4$ are three new functions of $\rho$. If
$D\not=2$, from the $\varepsilon^4$ equations also follows that
\begin{equation}
b_2=\frac{a_2}{2-D} \,,  \label{eq:b2}
\end{equation}
and that the functions $a_2$ and $p_2$ are determined by the coupled
differential equations
\begin{align}
\frac{d^2a_2}{d\rho^2}+\frac{D-1}{\rho}\,\frac{da_2}{d\rho}&=
\frac{D-2}{D-1}\,p_2^2 \,, \label{eq:x2}\\
\frac{d^2p_2}{d\rho^2}+\frac{D-1}{\rho}\,\frac{dp_2}{d\rho}&=
p_2(a_2-\omega_2) \,. \label{eq:p2}
\end{align}
If $D=2$ then $a_2=0$, and there are no nontrivial localized regular
solutions for $b_2$ and $p_2$, so we assume $D>2$ from now. We note
that at all orders $\sin\tau$ terms can be absorbed by a small shift
in the time coordinate. After this, no $\sin(k\tau)$ terms appear in
the expansion, resulting in the time reflection symmetry at $\tau=0$.

Since we have already set $m^2=u_1=v_1=1$, equations \eqref{eq:x2} and
\eqref{eq:p2} do not depend on the coefficients $v_k$ of the potential
$\bar U(\phi)$. To order $\varepsilon^2$ the functions $\phi$, $A$ and
$B$ are the same for any potential. This means that the leading order
small-amplitude behavior of oscillatons is always the same as for the
Klein-Gordon case.

\subsection{Schr\"odinger-Newton equations}\label{sec:SN}

Introducing the functions $s$ and $S$ by
\begin{equation}
s=\omega_2-a_2 \ , \quad S=p_2\sqrt{\frac{D-2}{D-1}} \ ,
\label{eq:sands}
\end{equation}
equations \eqref{eq:x2} and \eqref{eq:p2} can be written into the form
which is called the time-independent Schr\"odinger-Newton (SN)
equations in the literature \cite{Ruffini,Friedberg,Ferrell,Moroz,Tod}:
\begin{align}
\frac{d^2S}{d\rho^2}
+\frac{D-1}{\rho}\,\frac{d S}{d\rho}
+s S&=0
\,, \label{Seq}\\
\frac{d^2s}{d\rho^2}
+\frac{D-1}{\rho}\,\frac{d s}{d\rho}
+S^2&=0
\,. \label{seq}
\end{align}
Equations \eqref{Seq} and \eqref{seq} have the scaling invariance
\begin{equation}
(S(\rho),s(\rho)) \to
(\lambda^2S(\lambda\rho),\lambda^2s(\lambda\rho))
\,.\label{snscale}
\end{equation}
If $2<D<6$ the SN equations have a family of solutions with $S$
tending to zero exponentially as $\rho\to\infty$, and $s$ tending to a
constant $s_0<0$ as
\begin{equation}
s\approx s_0+s_1\rho^{2-D} \,. \label{eq:sasympt}
\end{equation}
The solutions are indexed by the number of nodes of $S$. The nodeless
solution corresponds to the lowest energy and most stable oscillaton.
We use the scaling freedom \eqref{snscale} to make the nodeless
solution unique by setting $s_0=\lim_{\rho\to\infty}s=-1$. At the same
time we change the $\varepsilon$ parametrization by requiring
\begin{equation}
\omega_2=-1 \ \  {\rm for} \ \  2<D<6 \,,
\end{equation}
ensuring that the limiting value of $a_2$ vanishes.
Then  for large $\rho$
\begin{equation}
a_2\approx-s_1\rho^{2-D} \ \  {\rm for} \ \  2<D<6 \,, \label{eq:a2asy}
\end{equation}
with only exponentially decaying corrections.  Going to higher orders,
it can be shown that one can always make the choice $\omega_i=0$ for
$i\geq 3$, thereby fixing the $\varepsilon$ parametrization, and
setting
\begin{equation}\label{eq:omeps}
\omega=\sqrt{1-\varepsilon^2} \ \  {\rm for} \ \  2<D<6 \,.
\end{equation}
For $D=6$ the explicit form of the asymptotically decaying solutions
are known
\begin{equation}
s=\pm S=\frac{24\alpha^2}{\left(1+\alpha^2\rho^2\right)^2}
 \ \  {\rm for} \ \  D=6 \,,
\label{6dss}
\end{equation}
where $\alpha$ is any constant.  In this case, since both $s$ and $S$
tend to zero at infinity, we have no method yet to fix the value of
$\alpha$ in \eqref{6dss}. Moreover, in order to ensure that $\varphi$
tends to zero at infinity we have to set
\begin{equation}
\omega_2=0 \ \  {\rm for} \ \  D=6 \,.
\end{equation}
For $D>6$ there are no solutions of the SN equations representing
localized configurations \cite{Choquard}.

Motivated by the asymptotic behavior of $s$, if $D\not=2$ it is
useful to introduce the variables
\begin{equation}
\sigma=\frac{\rho^{D-1}}{2-D}\,\frac{d s}{d\rho} \,, \qquad
\nu=s-\rho^{2-D}\sigma \ . \label{eq:sigmadef}
\end{equation}
In $2<D<6$ dimensions these variables tend exponentially to the
earlier introduced constants
\begin{equation}
\lim_{\rho\to\infty}\sigma=s_1 \,, \qquad
\lim_{\rho\to\infty}\nu=s_0 \,. \label{e:s0s1}
\end{equation}
Then the SN equations can be written into the equivalent form
\begin{align}
\frac{d\sigma}{d\rho}+\frac{\rho^{D-1}}{2-D}S^2&=0
\,,\label{e:dsigma}\\
\frac{d\nu}{d\rho}+\frac{\rho}{D-2}S^2&=0 \,,\\
\frac{d^2S}{d\rho^2}
+\frac{D-1}{\rho}\,\frac{d S}{d\rho}
+\left(\nu+\rho^{2-D}\sigma\right) S&=0 \,,
\end{align}
which is more appropriate for finding high precision numerical
solutions. Equation \eqref{e:dsigma} will turn out to be useful when
integrating the mass-energy density in Section \ref{sec:propermass} in
order to determine the proper mass.

\subsection{Higher order expansion}\label{sec:next}

From the $\varepsilon^6$ components of the field equations follows the
time dependence of $A_4$,
\begin{equation}
A_4=a_4^{(0)}+a_4^{(2)}\cos(2\tau) \,,
\end{equation}
where $a_4^{(0)}$ and $a_4^{(2)}$ are functions of $\rho$. The
functions $p_4$ and $a_4^{(0)}$ are determined by the coupled
equations
\begin{align}
&\frac{d^2a_4^{(0)}}{d\rho^2}+\frac{D-1}{\rho}\,\frac{da_4^{(0)}}{d\rho}=
\frac{2p_2p_4(D-2)}{D-1} \notag\\
&\qquad+\left(\frac{da_2}{d\rho}\right)^2
+\omega_2p_2^2-\frac{2p_2^2a_2}{D-1}
, \label{eq:a40}\\
&\frac{d^2p_4}{d\rho^2}+\frac{D-1}{\rho}\,\frac{dp_4}{d\rho}=
p_4(a_2-\omega_2)\notag\\
&\qquad+\left(a_4^{(0)}-\omega_4\right)p_2
-\frac{a_2p_2(D-1)(a_2-\omega_2)}{D-2}\label{eq:p4}\\
&\qquad-\frac{Dp_2^3}{8(D-1)}-\left(\frac{5}{6}v_2^2
-\frac{3}{4}v_3\right)p_2^3 \,. \notag
\end{align}
We look for the unique solution for which both $a_4^{(0)}$ and $p_4$
tend to zero as $\rho\to\infty$. For $2<D<6$ the function $p_4$ goes
to zero exponentially, while for large $\rho$
\begin{equation}
a_4^{(0)}\approx\frac{1}{2}s_1^2\rho^{4-2D}+s_2\rho^{2-D}+s_3
\label{eq:a4asympt}\,,
\end{equation}
where $s_1$ is defined in \eqref{eq:sasympt}, and $s_2$ and $s_3$ are
some constants. If $a_4^{(0)}$ and $p_4$ are solutions of
\eqref{eq:a40} and \eqref{eq:p4}, then for any constant $c$
\begin{align}
\bar a_4^{(0)}&=a_4^{(0)}+c\left[2(a_2-\omega_2)
+\rho\frac{da_2}{d\rho}\right] , \\
\bar p_4&=p_4+c\left(2p_2+\rho\frac{dp_2}{d\rho}\right) ,
\end{align}
are also solutions. This family of solutions is generated by the
scaling freedom \eqref{snscale} of the SN equations. If we have any
solution of \eqref{eq:a40} and \eqref{eq:p4} then by choosing $c$
appropriately we can get another solution for which $s_3=0$ in
\eqref{eq:a4asympt}.

The equation for $b_4$ is
\begin{align}
\frac{db_4}{d\rho}&=\frac{1}{2-D}\,\frac{da_4^{(0)}}{d\rho}\notag\\
&+\frac{1}{4(D-2)^2}\,\frac{da_2}{d\rho}
\left[\rho\frac{da_2}{d\rho}+4(D-1)a_2\right] \label{eq:sb4}\\
&+\frac{\rho}{2(D-1)(D-2)}
\left[\left(\frac{dp_2}{d\rho}\right)^2-p_2^2(a_2-\omega_2)\right]
. \notag
\end{align}
For large $\rho$ the function $b_4$ tends to zero as
\begin{equation}
b_4\approx \frac{6-D}{8(D-2)^2}s_1^2\rho^{4-2D}
+\frac{s_2}{2-D}\rho^{2-D} \,.
\end{equation}
The $\cos(2\tau)$ part of $A_4$ is determined by
\begin{align}
\frac{d^2a_4^{(2)}}{d\rho^2}-\frac{1}{\rho}\,\frac{da_4^{(2)}}{d\rho}&=
\frac{(D-2)(a_2-\omega_2)p_2^2}{2(D-1)}  \label{eq:sa42}\\
&-\frac{D}{2(D-1)}\,\frac{dp_2}{d\rho}
\left(\frac{dp_2}{d\rho}+\frac{D-2}{\rho}p_2\right) . \notag
\end{align}
We remind the reader, that for $2<D<6$ the choice $\omega_2=-1$,
$\omega_4=0$ is natural, while for $D=6$ necessarily $\omega_2=0$. For
a Klein-Gordon field in $D=6$ the only nonvanishing coefficient is
$\omega_4=-1$.

Summarizing the results, the scalar field and the metric components up
to $\varepsilon^4$ order are
\begin{align}
\phi&=\varepsilon^2p_2\cos\tau \label{eq:phisum}\\
&+\varepsilon^4\left\{
p_4\cos\tau+\frac{v_2p_2^2}{6}\left[\cos(2\tau)-3\right]\right\}
+\mathcal{O}(\varepsilon^6) \,, \notag\\
A&=1+\varepsilon^2a_2+\varepsilon^4\left[
a_4^{(0)}+a_4^{(2)}\cos(2\tau)\right]+\mathcal{O}(\varepsilon^6)
\,, \\
B&=1-\varepsilon^2\frac{a_2}{D-2}\label{eq:bsum}\\
&+\varepsilon^4\left[
b_4-\frac{p_2^2}{4(D-1)}\cos(2\tau)
\right]+\mathcal{O}(\varepsilon^6)
\,. \notag
\end{align}

Going to higher orders, the expressions get rather
complicated. However, it can be seen that for symmetric potentials,
when $v_{2k}=0$, the scalar field $\phi$ contains only $\cos(k\tau)$
components with odd $k$, while $A$ and $B$ only contains even Fourier
components.

Some of the higher order expressions simplifies considerably when
considering symmetric potentials with $v_{2k}=0$. Because the first
radiating mode proportional to $\cos(3\tau)$ emerges at
$\varepsilon^6$ order in $\phi$ in symmetric potentials, we present its higher order
expression for the symmetric case
\begin{align}
\phi&=\varepsilon^2p_2\cos\tau
+\varepsilon^4p_4\cos\tau+\varepsilon^6p_6\cos\tau \label{eq:phisum3}\\
&+\varepsilon^6\left(
\frac{Dp_2^3}{64(D-1)}
+\frac{v_3p_2^3}{32}+\frac{p_2a_4^{(2)}}{8}
\right)\cos(3\tau)
+\mathcal{O}(\varepsilon^8) \,, \notag
\end{align}
where $p_6$ is a function of $\rho$ determined by lengthy differential
equations arising at higher orders.

\begin{figure}[!ht]
    \begin{center}
    \includegraphics[width=8cm]{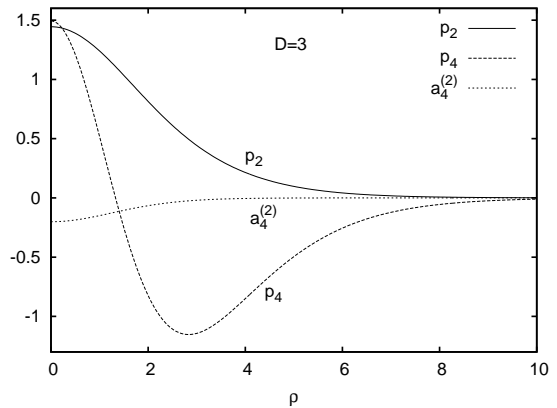}
    \end{center}
    \caption{The exponentially decaying functions $p_2$, $p_4$, and
      $a_4^{(2)}$ for the small-amplitude expansion of the
      Klein-Gordon oscillaton in the $D=3$ case.
      \label{f:exp}}
\end{figure}

\begin{figure}[!ht]
    \begin{center}
    \includegraphics[width=8cm]{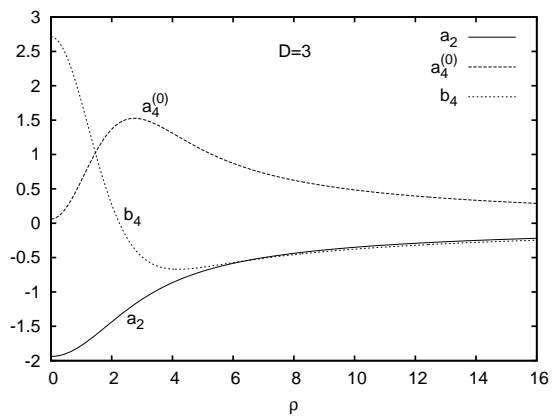}
    \end{center}
    \caption{The functions $a_2$, $a_4^{(0)}$, and $b_4$ for the $D=3$
      Klein-Gordon system. These functions tend to zero according to a
      power law for $\rho\to\infty$.
 \label{f:pow}}
\end{figure}

For the Klein-Gordon case in $D=3$ spatial dimensions we plot the
numerically obtained functions $p_2$, $a_2$, $p_4$, $a_4^{(0)}$,
$a_4^{(2)}$ and $b_4$ on Figs. \ref{f:exp} and \ref{f:pow}.

Equations \eqref{eq:phisum}-\eqref{eq:bsum} determine a one-parameter
family of oscillating configurations depending on the parameter
$\varepsilon$. This family solves the field equations with a scalar
field mass $m=1$. By applying the rescaling \eqref{eq:trans} to the
$t$ and $r$ coordinates, we can obtain one-parameter families of
solutions with any scalar mass $m$.

To $\varepsilon^2$ order, the metric is static. This is the biggest
advantage of the spatially conformally flat coordinate system $C=r^2B$
over the Schwarzschild area coordinates $C=r^2$. In the Schwarzschild
system the constant $r$ observers ``feel'' an $\varepsilon^2$ order
small oscillation in the metric (see Appendix \ref{app:schw}). The
magnitude of the acceleration of the constant $(r,\theta_1,\theta_2
...)$ observers in the general metric \eqref{eq:metrgen} is
\begin{equation}
\mathrm{a}=\frac{1}{2A\sqrt{B}}\,\frac{dA}{dr} \,, \label{eq:acc}
\end{equation}
which has an $\varepsilon^3$ order oscillating component when using
Schwarzschild coordinates, while in spatially conformally flat
coordinates the temporal change in the acceleration is only of order
$\varepsilon^5$.

The function $W=AB^{D-2}$ is equal to $1$ to order $\varepsilon^4$ in
the conformally flat coordinates. This motivates the metric form
choice
\begin{align}
ds^2&=-Adt^2 \\
&+\left(\frac{W}{A}\right)^{\frac{1}{D-2}}
\left(dr^2+r^2d\theta_1^2+r^2\sin^2\theta_1d\theta_2^2+\ldots\right)
 , \notag
\end{align}
which has been employed for the $D=3$ case in \cite{Page}.

\subsection{Singularities on the complex plane} \label{sec:sing}

As we will see in Section \ref{sec:radlaw}, in order to determine the
energy loss of oscillatons it is advantageous to extend the functions
$\phi$, $A$ and $B$ to the complex plane.  In the small-amplitude
expansion formalism the extension of the coefficient functions
$\phi_k$, $A_k$ and $B_k$ have symmetrically positioned poles along
the imaginary axis, induced by the poles of the SN equations. We
consider the closest pair of singularities, located at $\rho=\pm iQ_D$,
since these will provide the dominant contribution to the energy
loss. The numerically determined location of the pole for the spatial
dimensions where there is an exponentially localized core is
\begin{align}
Q_3&=3.97736 \,,\\
Q_4&=2.30468 \,,\\
Q_5&=1.23595 \,.
\end{align}

The leading order behavior of the functions near the poles can be
determined analytically, even if the solution of the SN equations is
only known numerically on the real axis.
Let us measure distances from the upper singularity by a coordinate
$R$ defined as
\begin{equation}
\rho=iQ_D+R\,.	 \label{eq:rrdef}
\end{equation}
Close to the pole we can expand the SN equations, and obtain that $s$
and $S$ have the same behavior,
\begin{equation}
s=S=-\frac{6}{R^2}-\frac{6i(D-1)}{5Q_DR}
-\frac{(D-1)(D-51)}{50Q_D^2}+{\cal O}(R) \ , \label{eq:sexp}
\end{equation}
even though they clearly differ on the real axis. We note that for
$D>1$ there are logarithmic terms in the expansion of $s$ and $S$,
starting with terms proportional to $R^4\ln R$.  According to
\eqref{eq:b2} and \eqref{eq:sands}, the expression \eqref{eq:sexp}
determines the $\varepsilon^2$ parts of $\phi$, $A$ and $B$ near the
pole.

Substituting into \eqref{eq:a40}, \eqref{eq:p4}, \eqref{eq:sb4} and
\eqref{eq:sa42}, the $\varepsilon^4$ order contributions $a_4^{(0)}$,
$p_4$, $b_4$ and $a_4^{(2)}$ can also be determined around the
pole. We give the results for the Klein-Gordon case, when $v_k=0$ for
$k>1$:
\begin{align}
&a_4^{(0)}=-\frac{9(25D+208)}{52(D-2)R^4}
+\frac{324iD(D-1)\ln R}{35Q_D(D-2)R^3}\notag\\
&\qquad+\frac{a_{-3}}{R^3}
+{\cal O}\left(\frac{\ln R}{R^2}\right) \,, \label{eq:a40r}\\
&p_4\sqrt{\frac{D-2}{D-1}}+a_4^{(0)}=
\frac{9(43D-104)}{26(D-2)R^4} \notag\\
&\qquad+\frac{9i(D-1)(3D-8)}{5Q_D(D-2)R^3}
+{\cal O}\left(\frac{1}{R^2}\right) \,,  \\
&b_4=\frac{9(333D+832)}{260(D-2)^2R^4}
-\frac{324iD(D-1)\ln R}{35Q_D(D-2)^2R^3} \\
&\qquad-\frac{a_{-3}}{(D-2)R^3}+
\frac{18i(D-1)}{5Q_D(D-2)R^3}
+{\cal O}\left(\frac{\ln R}{R^2}\right) \,, \notag\\
&a_4^{(2)}=-\frac{9(6-D)}{5(D-2)R^4}\notag\\
&\qquad+\frac{6i(D-1)(D-6)}{5Q_D(D-2)R^3}
+{\cal O}\left(\frac{1}{R^2}\right) \,. \label{eq:a42r}
\end{align}
The constant $a_{-3}$ can only be determined from the specific
behavior of the functions on the real axis, namely from the
requirement of the exponential decay of $p_4$ for large real $\rho$.

\section{Proper and total mass}\label{sec:masses}

\subsection{Proper mass} \label{sec:propermass}

In this subsection we present the calculation of the proper mass
$M_p$, which is usually obtained by the integral of the mass-energy
density over a spatial slice of the corresponding spacetime. In the
next subsection the calculation of the total mass $M$ will be
performed, by investigating the asymptotic behavior of the metric
components. The difference $E_b=M_p-M$ defines the gravitational
binding energy, which is expected to be positive.

The mass-energy density is $\mu=T_{ab}u^au^b$, where the unit timelike
vector $u^a$ has the components $(1/\sqrt{A},0,...,0)$.  In terms of
the rescaled scalar field $\phi$,
\begin{equation}
\mu=\frac{1}{8\pi}\left[\frac{1}{2A}\left(\frac{d\phi}{dt}\right)^2
+\frac{1}{2B}\left(\frac{d\phi}{dr}\right)^2+\bar U(\phi)\right]
\label{eq:massen}\,.
\end{equation}
The total proper mass in the metric \eqref{eq:metrgen} is defined by
the $D$ dimensional volume integral
\begin{equation}
M_p=\frac{2\pi^{\frac{D}{2}}}{\Gamma\left(\frac{D}{2}\right)}
\int_0^\infty dr\mu\sqrt{BC^{D-1}} \,. \label{eq:massint}
\end{equation}
Applying this for the small-amplitude expansion of oscillatons in
spatially conformally flat coordinates, and using that
$\rho=\varepsilon r$ and $\omega^2=1-\varepsilon^2$, we can write
\begin{align}
M_p&=\frac{2\pi^{\frac{D}{2}}}{8\pi\Gamma\left(\frac{D}{2}\right)}
\int_0^\infty d\rho\rho^{D-1}\varepsilon^{-D}\left(
1+\varepsilon^2b_2\right)^{\frac{D}{2}} \times\notag\\
&\Biggl\{
\frac{1}{2(1+\varepsilon^2a_2)}\biggl(\varepsilon^2p_2\omega\sin\tau
+\varepsilon^4p_4\sin\tau \notag\\
&+\varepsilon^4\frac{v_2p_2^2}{3}
\sin(2\tau)\biggr)^2
+\frac12\left(\varepsilon^3\cos\tau\frac{dp_2}{d\rho}\right)^2 \\
&+\frac12\biggl[\varepsilon^2p_2\cos\tau
+\varepsilon^4p_4\cos\tau \notag\\
&+\varepsilon^4\frac{v_2p_2^2}{3}
\left(\cos(2\tau)-3\right)\biggr]^2
+\frac{v_2}{3}\left(\varepsilon^2p_2\cos\tau\right)^3
\Biggr\} . \notag
\end{align}
Using \eqref{eq:b2} and \eqref{eq:p2}, for the proper mass we obtain
\begin{equation}
M_p=\varepsilon^{4-D}M_p^{(1)}
+\varepsilon^{6-D}M_p^{(2)}
+\mathcal{O}\left(\varepsilon^{8-D}\right) ,
\end{equation}
where
\begin{align}
M_p^{(1)}&=\frac{\pi^{\frac{D}{2}}}
{8\pi\Gamma\left(\frac{D}{2}\right)}
\int_0^\infty d\rho\rho^{D-1}p_2^2 \,, \label{eq:mp1}\\
M_p^{(2)}&=\frac{\pi^{\frac{D}{2}}}
{8\pi\Gamma\left(\frac{D}{2}\right)}
\int_0^\infty d\rho\rho^{D-1}
\biggl(2p_2p_4\notag\\
&\qquad\qquad\qquad-p_2^2-\frac{3D-4}{2(D-2)}a_2p_2^2\biggr)
\,. \label{eq:mp2}
\end{align}
The result turns out to be time independent to this order.  Although
the coefficients $v_i$ of the potential also drop out from
\eqref{eq:mp2}, the dependence on the form of the potential still
comes in through \eqref{eq:p4}. The leading order behavior,
\eqref{eq:mp1}, only depends on the scalar field mass $m$, which has
been rescaled to $1$ for simplicity. Applying \eqref{eq:trans} to
obtain solutions with $m\not=1$, an $m^2$ factor appears in
\eqref{eq:massen} for the mass-energy density $\mu$. Since the volume
element in the integral contains a $m^{-D}$ factor, in all the
presented proper and total mass formulas an $m^{2-D}$ factor appears.

Using \eqref{eq:sands}, \eqref{e:s0s1} and \eqref{e:dsigma}, the
leading order coefficient is
\begin{equation}
M_p^{(1)}=\frac{(D-1)\pi^{\frac{D}{2}}}
{8\pi\Gamma\left(\frac{D}{2}\right)}\,s_1 \,. \label{eq:propm1}
\end{equation}
The numerically calculated values for $M_p^{(1)}$ and $M_p^{(2)}$ for
the Klein-Gordon field case in various spatial dimensions are listed
in Table \ref{masstable}.
\begin{table}[htbp]
\begin{tabular}{|c|c|c|c|}
\hline
  & $D=3$  & $D=4$  & $D=5$ \\
\hline
$M^{(1)}=M_p^{(1)}$ &  $1.75266$ &   $9.06533$ &  $21.7897$ \\
$M^{(2)}$          & $-2.11742$ & $-43.5347$ & $-533.732$ \\
$M_p^{(2)}$        & $-1.53319$ & $-39.0020$ & $-555.521$ \\
\hline
\end{tabular}
\caption{\label{masstable}
  Coefficients of the $\varepsilon$ expansion of the total mass $M$
  and proper mass $M_p$ for the $m=1$ Klein-Gordon case in $D=3,4,5$
  spatial dimensions.
}
\end{table}

\subsection{Total mass} \label{sec:totmass}

Since the scalar field tends to zero exponentially, at large distances
the metric should approach the static Schwarzschild-Tangherlini
metric \cite{Tangherlini}.  In Schwarzschild area coordinates, with
$C=r^2$, this metric has the form
\begin{equation}
ds^2=-\left(1-\frac{r_0^{D-2}}{r^{D-2}}\right)dt^2
+\frac{1}{1-\frac{\textstyle r_0^{D-2}}{\textstyle r^{D-2}}}\,dr^2
+r^2d\Omega_{D-1}^2 \,,
\end{equation}
while in the spatially conformally flat coordinate system, $C=r^2B$,
it can be written as
\begin{align}
ds^2&=-\left(\frac{4r^{D-2}-r_0^{D-2}}
{4r^{D-2}+r_0^{D-2}}\right)^2dt^2
\notag\\
&+\left(1+\frac{r_0^{D-2}}{4r^{D-2}}\right)^{\frac{4}{D-2}}\left(dr^2
+r^2d\Omega_{D-1}^2\right) \,,
\end{align}
where $r_0$ is a constant related to the mass.  In general spherically
symmetric spacetimes it is possible to define the natural radius
function $\hat r$ by \eqref{eq:radfunc}, and the mass function $\hat
m$ by \eqref{eq:massfunc}.  In both the Schwarzschild and conformally
flat coordinates, for the Schwarzschild-Tangherlini metric $\hat m$ is
constant,
\begin{equation}
\hat m=M=\frac{(D-1)\pi^{\frac{D}{2}}}
{8\pi\Gamma\left(\frac{D}{2}\right)}
r_0^{D-2}\,.
\end{equation}

For the small-amplitude expansion of oscillatons in the spatially
conformally flat coordinate system the radius function is $\hat
r=r\sqrt{B}$, where $B$ is expanded according to
\eqref{eq:bexp}. Using the rescaled radial coordinate
$\rho=\varepsilon r$, the mass function can be expanded as
\begin{equation}
\hat m=\varepsilon^{4-D}\hat m^{(1)}
+\varepsilon^{6-D}\hat m^{(2)}
+\mathcal{O}\left(\varepsilon^{8-D}\right) ,
\end{equation}
where
\begin{equation}
\hat m^{(1)}=-\frac{(D-1)\pi^{\frac{D}{2}}}
{8\pi\Gamma\left(\frac{D}{2}\right)}\rho^{D-1}\frac{dB_2}{d\rho} \,,
\end{equation}
and
\begin{align}
\hat m^{(2)}&=-\frac{(D-1)\pi^{\frac{D}{2}}}
{8\pi\Gamma\left(\frac{D}{2}\right)}\rho^{D-1}
\Biggl[\frac{dB_4}{d\rho}\notag\\
&\qquad+\frac{\rho}{4}\left(\frac{dB_2}{d\rho}\right)^2
+\frac{D-4}{2}B_2\frac{dB_2}{d\rho}\Biggr] \,.
\end{align}
The total mass is the limit at $r\to\infty$,
\begin{equation}
M=\varepsilon^{4-D}M^{(1)}
+\varepsilon^{6-D}M^{(2)}
+\mathcal{O}\left(\varepsilon^{8-D}\right) . \label{eq:totmass}
\end{equation}
Since $B_2=a_2/(2-D)$, using the asymptotic form \eqref{eq:a2asy} of
$a_2$, we get
\begin{equation}
M^{(1)}=\frac{(D-1)\pi^{\frac{D}{2}}}
{8\pi\Gamma\left(\frac{D}{2}\right)}\,s_1 \,,
\end{equation}
agreeing with the leading order coefficient of the proper mass,
$M_p^{(1)}$, given in \eqref{eq:propm1}. Using \eqref{eq:sb4} for the
derivative of $b_4$,
\begin{align}
M^{(2)}&=\lim_{\rho\to\infty}\frac{(D-1)\pi^{\frac{D}{2}}}
{8\pi\Gamma\left(\frac{D}{2}\right)}\,\frac{\rho^{D-1}}{D-2}
\Biggl[\frac{da_4^{(0)}}{d\rho} \notag\\
&\qquad-\frac{\rho}{2(D-2)}\left(
\frac{da_2}{d\rho}\right)^2-\frac{3}{2}a_2\frac{da_2}{d\rho}\Biggr] ,
\end{align}
which can be easily calculated numerically, since the expression in
the limit tends to a constant exponentially. The numerical results for
the Klein-Gordon field are presented in Table \ref{masstable}. The
proper mass and the total mass agree to leading order. However, taking
into account the next term in the expansion, it turns out that, as it
can be expected, the gravitational binding energy $E_b=M_p-M$ is
positive.

It is instructive to write the expressions for the total mass in
natural units, where the oscillaton mass $M$ is measured in kilograms
while the mass of the scalar field $m$ in units of $eV/c^2$:
\begin{align}
M_{(D=3)}&=\varepsilon\left(4.66-5.63\,\varepsilon^2\right)
10^{20}kg\frac{eV}{mc^2} \,, \\
M_{(D=4)}&=\left(2.94-14.1\,\varepsilon^2\right)
10^{49}kg\left(\frac{eV}{mc^2}\right)^2 \,, \\
M_{(D=5)}&=\frac{1}{\varepsilon}
\left(8.63-211\,\varepsilon^2\right)
10^{77}kg\left(\frac{eV}{mc^2}\right)^3 \,.
\end{align}
Since $\mathcal{O}\left(\varepsilon^{4}\right)$ terms were dropped,
these expressions are precise only for small values of
$\varepsilon$. However, comparing to the $3+1$ dimensional numerical
results obtained by solving the Fourier mode equations in
\cite{LopezMatos}, it can be inferred that these total mass
expressions give a reasonable estimate even when $\varepsilon\approx
0.5$.

\subsection{Size of oscillatons}

Although oscillatons are exponentially localized, they do not have a
definite outer surface. A natural definition for their size is to take
the radius $r_n$ inside which $n$ percentage of the mass can be
found. It is usual to take, for example, $n=95$. The mass inside a
given radius $r$ can be defined either by the integral
\eqref{eq:massint} replacing the upper limit by $r$, or by taking the
local mass function $\hat m$ in \eqref{eq:massfunc}. To leading order
in $\varepsilon$ both definitions give
\begin{equation}
M(r)=\frac{(D-1)\pi^{\frac{D}{2}}}
{8\pi\Gamma\left(\frac{D}{2}\right)}\,\sigma(\varepsilon r) \,,
\end{equation}
where $\sigma$ has been introduced in \eqref{eq:sigmadef} as a
function of $\rho=\varepsilon r$. The rescaled radius $\rho_{n}$ can
be defined by
\begin{equation}
\frac{\sigma(\rho_{n})}{\sigma(\infty)}=\frac{n}{100} \,.
\end{equation}
The numerical values of $\rho_{n}$ for various $n$ in $D=3,4,5$
dimensions are listed in Table \ref{tablerhon}.
\begin{table}[htbp]
\begin{tabular}{|c|c|c|c|}
\hline
  & $D=3$  & $D=4$  & $D=5$ \\
\hline
$\rho_{50}$   &  $2.240$ &   $1.778$ &  $1.317$ \\
$\rho_{90}$   &  $3.900$ &   $3.013$ &  $2.284$ \\
$\rho_{95}$   &  $4.471$ &   $3.455$ &  $2.652$ \\
$\rho_{99}$   &  $5.675$ &   $4.410$ &  $3.478$ \\
$\rho_{99.9}$ &  $7.239$ &   $5.692$ &  $4.634$ \\
\hline
\end{tabular}
\caption{\label{tablerhon}
  The radius inside which given percentage of the mass is contained for
  various spatial dimensions.
}
\end{table}
Restoring the scalar field mass $m$ into the expression, the physical
radius is
\begin{equation}
r_{n}=\frac{\rho_{n}}{\varepsilon m} \,.
\end{equation}
In natural units, measuring $mc^2$ in electron volts and $r_n$ in
meters (Roman $\mathrm{m}$),
\begin{equation}
r_{n}=1.97\cdot 10^{-7}\mathrm{m}\,\frac{\rho_{n}}{\varepsilon}
\,\frac{eV}{mc^2} \,. \label{eq:size}
\end{equation}
Similarly to the total mass expressions in the previous subsection,
this result is still a reasonable approximation for as large
$\varepsilon$ values as $0.5$.

\subsection{Stability}\label{sec:stab}

The stability properties of oscillatons are in many respects very
similar to cold neutron and boson stars.  Perfect fluid stars are
known to be stable for small $\mu_c$ central densities. As $\mu_c$
increases, the total mass $M$ also increases, until it reaches a
maximal value $M_{\mathrm{max}}$, where according to a theorem in
\cite{Harrison}, an unstable radial mode sets in. Boson stars have
analogous stability properties \cite{leepang}.  For oscillatons in the
EKG system (for $D=3$) this behavior has also been observed in Refs.\
\cite{Seidel1} and \cite{Alcub}.  Oscillatons are closely related to
flat background oscillons, which also behave very similarly to neutron
and boson stars. Since we use a small-amplitude expansion for
oscillons and oscillatons it is more instructive to use the magnitude
of the oscillating central amplitude $\Phi_c$, instead of the central
density $\mu_c$. The central density is expected to be a monotonically
increasing function of the central amplitude.

Given the very close analogy with oscillons we formulate
a general conjecture on the stability of oscillatons.
We recall that in all known examples
the stability pattern of oscillons is the same, namely if
$dE/d\varepsilon>0$ oscillons are stable, while when
$dE/d\varepsilon<0$ oscillons are unstable, where $E=E(\varepsilon)$
is the total energy of the oscillon \cite{FFHL,FFHM,moredim,dilaton}.
Therefore we conjecture that the same stability pattern holds true for
oscillatons, except that the energy, $E$ is replaced by the total mass
$M=M(\varepsilon)$ of the oscillaton.
In other words if the time evolution (i.e. energy/mass
loss) of an oscillon/oscillaton leads to spreading of the core, the
oscillon/oscillaton is stable, while oscillons/oscillatons are
unstable if they have to contract with time evolution. Therefore if
the conjecture is true, the first two terms in the expansion of $M$
enables us to determine the stability of oscillatons.

Taking into account the first two terms in \eqref{eq:totmass}, for the
$D=3$ Klein-Gordon field the total mass has a maximum at
\begin{equation}
\varepsilon_{\mathrm{max}}=\sqrt{\frac{-M^{(1)}}{3M^{(2)}}}
\approx 0.525 \,,
\end{equation}
corresponding to the value of the frequency, $\omega_{\mathrm{min}}\approx
0.851$. Although this is just a leading order result, it agrees reasonably well
with the frequency $0.864$ obtained by the
numerical solution of the Fourier mode equations in
\cite{LopezMatos}. The value of the mass at the maximum is
\begin{equation}
M_{\mathrm{max}}=\frac{2}{3}\varepsilon_{\mathrm{max}}M^{(1)}
\approx 0.614 \,,
\end{equation}
which is also quite close to the number $0.607$ given in
\cite{LopezMatos}.

For an axion with $m=10^{-5}eV/c^2$ the maximal
mass is $M_{\mathrm{max}}=1.63\cdot 10^{25}kg$, which is about three times the mass of
the Earth. The radius of this oscillaton, according to the leading
order approximation \eqref{eq:size}, is $r_{95}=16.8\,\mathrm{cm}$,
while its Schwarzschild radius is $2.42\,\mathrm{cm}$.

The mass maximum is very important concerning the stability of
oscillatons. According to \cite{Seidel1,Alcub}, for the Klein-Gordon
field in $D=3$ small-amplitude oscillatons with
$\varepsilon<\varepsilon_{\mathrm{max}}$ are stable, while those with
$\varepsilon>\varepsilon_{\mathrm{max}}$ are unstable, presumably
having a single decay mode. For $D=4$ and $D=5$ the mass is a
monotonically decreasing function of $\varepsilon$, and all
oscillatons are expected to be unstable. The situation may be totally
different for scalar fields with a nontrivial potential $U(\phi)$. For
example, for $v_2=0$ and $v_3=4$, the coefficient $M^{(2)}$ becomes
positive, and there is no maximum on the energy curve in $D=3$
dimensions, consequently even large amplitude oscillatons can be
expected to be stable. Using this potential in $D=4$ dimensions, for
small $\varepsilon$ the mass will be a monotonically increasing
function, so small-amplitude configurations should be stable. For
$D=5$, in this case, there is a minimum in the energy curve, above
which stable oscillatons can be expected.

\section{Radiation law of oscillatons} \label{sec:radlaw}

The methods that we apply in this section for the calculation of the
radiation law of oscillatons have been already applied for oscillons
formed by scalar fields on flat background. The extension of the
Fourier mode equations to the complex plane has been first used for
the one-dimensional $\phi^4$ theory by Segur and Kruskal
\cite{SK}. The Borel summation method to calculate the small
correction near the pole has been introduced by Pomeau, Ramani and
Grammaticos \cite{Pomeau}. The results has been extended to higher
dimensional oscillons in \cite{moredim} and to a scalar-dilaton system
in \cite{dilaton}.

\subsection{Fourier expansion}

Since all terms in the expansion \eqref{eq:phiexp}-\eqref{eq:bexp} are
exponentially decaying, the small-amplitude expansion can only be
applied to the core region of oscillatons. It cannot describe the
exponentially small radiative tail responsible for the energy
loss. This is closely related to the fact that the expansion is not
convergent, it is an asymptotic expansion. Instead of studying a
radiating oscillaton configuration with slowly varying frequency, it
is simpler to consider exactly periodic solutions having a relatively
large amplitude core and a very small amplitude standing wave tail.
We Fourier expand the scalar and the metric components as
\begin{align}
\phi&=\sum_{k=0}^{N_F}\bar\phi_{k}\cos(k\omega t)
\,, \label{eq:fourexp1}\\
A&=1+\sum_{k=0}^{N_F}\bar A_{k}\cos(k\omega t)
\,, \\
B&=1+\sum_{k=0}^{N_F}\bar B_{k}\cos(k\omega t) \,, \label{eq:fourexp3}
\end{align}
where $\bar\phi_{k}$ $\bar A_{k}$ and $\bar B_{k}$ only depend on $r$,
and solve the Fourier mode equations obtained from Einstein's
equations and the wave equation. Although, in principle, the Fourier
truncation order $N_F$ should tend to infinity, one can expect very
good approximation for moderate values of $N_F$. We assume that the
frequency is approaching from below the mass threshold $m=1$, and in
this context, define the $\varepsilon$ parameter by
$\varepsilon=\sqrt{1-\omega^2}$.

Regularity at the center require finite values for $\bar\phi_k$, $\bar
A_k$ and $\bar B_k$ for $r=0$, together with
\begin{equation}
\left.\frac{d\bar\phi_k}{dr}\right|_{r=0}=0 \,, \quad
\left.\frac{d\bar A_k}{dr}\right|_{r=0}=0 \,, \quad
\left.\frac{d\bar B_k}{dr}\right|_{r=0}=0 \,.
\end{equation}
Concerning the boundary conditions at $r\to\infty$, it is a natural
but quite restrictive requirement to assume that the metric is
asymptotically flat, with the $t$ coordinate tending to the proper
time for large radii. This implies that $\bar A_k\to 0$ and $\bar
B_k\to 0$ for $r\to\infty$. The Fourier components of the wave
equation \eqref{eq:wave3} for large $r$ decouple, and in this case can
be written as
\begin{equation}
  \frac{d^2\bar\phi_n}{dr^2}
  +\frac{D-1}{r}\,\frac{d\bar\phi_n}{dr}
  +(n^2\omega^2-1)\bar\phi_n=0 \,. \label{eq:larger}
\end{equation}
In the relevant frequency range $1/2<\omega<1$, if $n\geq2$ these
equations have oscillatory solutions, behaving asymptotically as
\begin{align}
\bar\phi_n&=
\frac{\gamma_n^{(s)}}{r^{(D-1)/2}}
\sin\left(r\sqrt{n^2\omega^2-1}\right) \\
&\quad+\frac{\gamma_n^{(c)}}{r^{(D-1)/2}}
\cos\left(r\sqrt{n^2\omega^2-1}\right) , \notag
\end{align}
where $\gamma_n^{(s)}$ and $\gamma_n^{(c)}$ are some constants.
According to \eqref{eq:massen}, this oscillating tail has a mass-energy
density, $\mu$, proportional to $r^{1-D}$. This implies that if
$\gamma_n^{(s)}$ or $\gamma_n^{(c)}$ is nonzero for any $n\geq2$, the
total proper mass of the spacetime is infinite. The requirement of the
vanishing of all these coefficients together with the central boundary
conditions are clearly too many conditions to satisfy for the given
number of second order differential equations. In general, regular
finite mass exactly periodic solutions are not expected to exist.

If we require that $\gamma_n^{(s)}=0$ and $\gamma_n^{(c)}=0$ for all
$n$ then all $\phi_n$ tend to zero exponentially, and we have a finite
mass asymptotically flat configuration. However, in general, this
solution will be singular at the center, hence we name this solution
\emph{singular breather} (SB). For a given frequency, $\omega$, the
singular breather solution is unique by parameter counting.

Because of their close similarity to oscillatons, it is important to
study another, presumably unique periodic solution, the so called
\emph{quasibreather} (QB) solution, which is regular at the center,
but which has a minimal energy density standing wave tail. Our aim is
to construct the quasibreather solution from the singular breather
solution. It is important to point out that the quasibreather picture
is only valid inside some large but finite radius. However
small the energy density of the oscillating tail is, going to very
large distances its contribution to the mass will not be negligible
anymore. Consequently, the assumption that $\bar A_k$ and $\bar B_k$
tends to zero will not remain true for arbitrarily large values
of $r$, and consequently Eq.\ \eqref{eq:larger} will also
change. For sufficiently large values of $r$,
the metric function $A$ increases until it causes to change the first radiating mode
(either $\phi_2$ or $\phi_3$) from oscillating to exponentially decaying.
Increasing $r$ further, all modes will stop oscillating one by one.
This way we obtain the exactly time-periodic
but infinite mass ``breathers'' which are described in details in
Section IX of the paper of Don N.~Page \cite{Page}. The quasibreather
can be considered to be the part of such an infinite mass "breather"
containing the core and a large portion of the tail where the first
radiating mode oscillates, requiring that the mass inside this region
is dominated by that of the oscillon core. Since the core amplitude is
of the order $\varepsilon^2$, while the tail is exponentially
suppressed in $\varepsilon$, the quasibreather picture is valid in a
sufficiently large volume.

Since the amplitude of the oscillating tail of the quasibreather is
very small, apart from a small region around the center the core of
the QB is very close to the corresponding singular breather solution.
In particular, the SB and the QB solutions have the same $\varepsilon$
expansions.  For the SB solution the small-amplitude expansion will
not be valid in a region near the center $r=0$, while for the QB
solution it will fail for large radii where the oscillating tail
becomes dominant. The size of the region, $(0,r_{\rm diff})$, around
$r=0$ where the difference between the SB and the regular core becomes
relevant is $r_{\rm diff}={\cal{O}}(e^{-\delta/\varepsilon})$
(with $\delta$ being a constant), whereas the size of the SB
or QB core is proportional to $1/\varepsilon$.  Outside of this
region, i.e.\ for $r>r_{\rm diff}$ the difference between the SB and
the QB will be very small since the singular mode turns out to be
proportional to the tail amplitude, which is exponentially small in
terms of the small parameter $\varepsilon$, while the core amplitude
is of order $\varepsilon^2$.

For potentials $U(\phi)$ which are symmetric around their minima,
i.e.\ $v_{2k}=0$ for integer $k$, the Fourier expansion of the scalar
contains only odd, while that of the metric components only even
terms,
\begin{equation}
\bar \phi_{2k}=0 \,, \quad \bar A_{2k+1}=0 \,, \quad
\bar B_{2k+1}=0 \,.
\end{equation}
For symmetric potentials the first radiating mode is $\bar \phi_{3}$.
In this section we will concentrate mainly on the Klein-Gordon scalar
field with $v_k=0$ for $k>1$. It is straightforward to generalize the
results for symmetric potentials.

For small-amplitude quasibreather or singular breather configurations
we can establish the connection between the Fourier expansion
\eqref{eq:fourexp1}-\eqref{eq:fourexp3} and the small-amplitude
expansion \eqref{eq:phiexp}-\eqref{eq:bexp} by comparing to
\eqref{eq:phisum}-\eqref{eq:phisum3}. For symmetric potentials we obtain:
\begin{align}
\bar\phi_1&=\varepsilon^2p_2+\varepsilon^4p_4
+\mathcal{O}(\varepsilon^6) \,, \label{eq:bphi1}\\
\bar\phi_3&=\varepsilon^6\left(\frac{Dp_2^3}{64(D-1)}
+\frac{v_3p_2^3}{32}+\frac{p_2a_4^{(2)}}{8}\right)
+\mathcal{O}(\varepsilon^8) \,,\\
\bar A_0&=\varepsilon^2a_2+\varepsilon^4a_4^{(0)}
+\mathcal{O}(\varepsilon^6) \,,\\
\bar A_2&=\varepsilon^4a_4^{(2)}
+\mathcal{O}(\varepsilon^6) \,,\\
\bar B_0&=-\varepsilon^2\frac{a_2}{D-2}+\varepsilon^4b_4
+\mathcal{O}(\varepsilon^6) \,,\\
\bar B_2&=-\varepsilon^4\frac{p_2^2}{4(D-1)}
+\mathcal{O}(\varepsilon^6) \,. \label{eq:bb2}
\end{align}

\subsection{Expansion near the pole}

As $\varepsilon\to 0$ the amplitude of all Fourier coefficients tend
to zero. However, extending them to the complex plane, for small
$\varepsilon$ they all have pole singularities on the imaginary axis
at $r=\pm iQ_D/\varepsilon$, corresponding to the poles of the
Schr\"odinger-Newton equations at $\rho=\pm iQ_D$, as it was discussed
in Sec.~\ref{sec:sing}. As $\varepsilon$ tends to zero, the poles move
further and further away from the real axis, but close to them the
Fourier components $\bar\phi_{k}$, $\bar A_k$ and $\bar B_{k}$ are not
getting small, in fact they have $\varepsilon$ independent parts. We
introduce a shifted radial coordinate, $y$, for an ``inner region'' around
the upper pole by
\begin{equation}
r=\frac{iQ_D}{\varepsilon}+y \,. \label{eq:ry}
\end{equation}
The coordinate $y$ is related to the one ($R$) defined in Eq.\ \eqref{eq:rrdef}
by $R=\varepsilon y$. Substituting the small-amplitude expansion
results \eqref{eq:sexp}-\eqref{eq:a42r} into
\eqref{eq:bphi1}-\eqref{eq:bb2}, and taking the limit $\varepsilon\to
0$, it follows that in the Klein-Gordon case, near the upper pole
\begin{align}
\bar\phi_1&=\left(-\frac{6}{y^2}+\frac{999D}{52(D-2)y^4}+...\right)
\sqrt{\frac{D-1}{D-2}} \,, \label{eq:phi1y}\\
\bar\phi_3&=\left(-\frac{27(7D-12)}{40(D-2)y^4}+...\right)
\sqrt{\frac{D-1}{D-2}} \,, \\
\bar A_0&=\frac{6}{y^2}-\frac{9(25D+208)}{52(D-2)y^4}+... \,, \\
\bar A_2&=-\frac{9(6-D)}{5(D-2)y^4}+... \,, \\
\bar B_0&=-\frac{6}{(D-2)y^2}
+\frac{9(333D+832)}{260(D-2)^2y^4}+... \,, \\
\bar B_2&=-\frac{9}{(D-2)y^4}+... \,. \label{eq:b1y}
\end{align}
We note that since \eqref{eq:phi1y}-\eqref{eq:b1y} are expansions in
$1/y^2$, they are valid for large $y$ values. In contrast,
\eqref{eq:sexp}-\eqref{eq:a42r} were calculated assuming small
$R$. Both of these conditions can hold simultaneously, since
$R=\varepsilon y$.

Expressions \eqref{eq:phi1y}-\eqref{eq:b1y} can also be obtained by
looking for the solution of the Fourier mode equations in the
$\varepsilon\to0$ limit near the pole as a power series expansion in
$1/y^2$,
\begin{align}
\bar\phi_{2k+1}&=\sum_{j=k+1}^{\infty}\psi_{2k+1}^{(j)}\frac{1}{y^{2j}}
\ , \label{eq:yexp1}\\
\bar A_{2k}&=\sum_{j=k+1}^{\infty}\alpha_{2k}^{(j)}\frac{1}{y^{2j}}
\ , \\
\bar B_{2k}&=\sum_{j=k+1}^{\infty}\beta_{2k}^{(j)}\frac{1}{y^{2j}} \ ,
\label{eq:yexp3}
\end{align}
where $\psi_{2k+1}^{(j)}$, $\alpha_{2k}^{(j)}$ and $\beta_{2k}^{(j)}$
are constants. The mode equations that we have to solve can be
obtained from the Einstein equations \eqref{eq:eieq1}-\eqref{eq:eieq4}
and from the wave equation \eqref{eq:wave3} by substituting
\eqref{eq:fourexp1}-\eqref{eq:fourexp3}. Equations
\eqref{eq:eieq1}-\eqref{eq:wave3} are not independent.  The wave
equation follows from the Einstein equations by the contracted Bianchi
identities, and the $(t,r)$ component \eqref{eq:eieq3} is a
constraint. The truncation of the Fourier expansion at a finite $N_F$
order makes the mode equations mutually contradictory. However, if we
choose any three field equations from
\eqref{eq:eieq1}-\eqref{eq:wave3}, the arising mode equations will
clearly have solutions. We have checked that our results for the
energy loss rate of oscillatons are the same for different choices of
the three field equations. We have also tested that the violation of
the mode equations obtained from the other two field equation tends to
zero quickly as $N_F$ increases.

Substituting \eqref{eq:ry} into the field equations
\eqref{eq:eieq1}-\eqref{eq:wave3} and taking the $\varepsilon\to0$
limit close to the pole, some terms with lower powers of $r$ can be
neglected. Then, inserting the $1/y^2$ expansion
\eqref{eq:yexp1}-\eqref{eq:yexp3} into the resulting mode equations,
because of the omission of odd powers of $1/y$, the only ambiguity
arises at the choice of the signature of $\psi_1^{(1)}$. The
calculation of \eqref{eq:phi1y}-\eqref{eq:b1y} using the Fourier mode
equations is technically more simple than using the small-amplitude
expansion method, and can be done by algebraic manipulation programs
to quite high orders in $1/y$.

Apart from an overall factor, the leading order behavior of the
coefficients $\psi_{k}^{(n)}$, $\alpha_{k}^{(n)}$ and
$\beta_{k}^{(n)}$ for large $n$ can be obtained by studying the
structure of the mode equations. It turns out that for large $n$,
$\psi_{3}^{(n)}$ dominates among the coefficients. For the third
Fourier mode of the Klein-Gordon field,
\begin{align}
\psi_{3}^{(n)}&=k_D(-1)^n\frac{(2n-1)!}{8^n}
\Biggl[1+\frac{3(9D-10)}{2(D-2)n} \notag\\
&\qquad+\frac{3(9D-10)(7D-8)}{2(D-2)^2n^2}
+{\cal O}\left(\frac{1}{n^3}\right)\Biggr] , \label{eq:p3nser}
\end{align}
where $k_D$ is a factor depending on $D$ and $N_F$. All other
coefficients grow slower with $n$ asymptotically. Although the $1/n$
and $1/n^2$ correction terms may depend on the choice of the scalar
potential, the leading order behavior is the same as in
\eqref{eq:p3nser} for any symmetric potential. The value of the
constant $k_D$ will turn out to be crucial for the determination of
the energy loss rate of oscillatons.  Calculating the coefficients up
to order $n=100$ and taking into account Fourier modes up to order
$N_F=6$, in the Klein-Gordon case we obtain
\begin{align}
k_3&=-0.301 \,, \\
k_4&=-0.134 \,, \\
k_5&=-0.0839 \,.
\end{align}

\subsection{The singular breather solution near the pole}

Expansion \eqref{eq:yexp1}-\eqref{eq:yexp3} gives an asymptotic series
representation of the Fourier components $\bar\phi_k$, $\bar A_k$ and
$\bar B_k$.  The results \eqref{eq:phi1y}-\eqref{eq:b1y} can be
considered as boundary conditions for the Fourier mode equations for
\begin{equation}
|y|\to\infty \ , \quad -\pi/2<\arg\, y<0 \,, \label{eq:matchreg}
\end{equation}
ensuring a unique solution for the ``inner problem''. This corresponds
to the requirement that $\phi$ decays to zero without any oscillating
tail for $r\to\infty$ along the positive half of the real axis,
i.e. we consider a singular breather solution.

The Fourier components of the wave equation \eqref{eq:wave3} can be
written as
\begin{equation}
\frac{d^2\bar\phi_n}{dr^2}
+\frac{D-1}{r}\,\frac{d\bar\phi_n}{dr}
+(n^2\omega^2-1)\bar\phi_n=F_n \,,
\label{eq:fourphi}
\end{equation}
where $F_n$ contain nonlinear polynomial terms in $\bar\phi_k$, $\bar
A_k$, $\bar B_k$ and their derivatives for $k\leq N_F$. Using the $y$
coordinate near the pole and taking the $\varepsilon\to 0$ limit,
\begin{equation}
\frac{d^2\bar\phi_n}{dy^2}
+(n^2-1)\bar\phi_n=\tilde F_n \,,
\label{eq:fourphipole}
\end{equation}
where $\tilde F_n$ denotes the $\varepsilon\to 0$ limit of $F_n$.  On
the imaginary axis the $1/y^2$ expansion gives real valued functions
to all orders. As singular breather solutions of the mode equations,
with boundary conditions \eqref{eq:phi1y}-\eqref{eq:b1y} in the region
given by \eqref{eq:matchreg}, the functions $\bar\phi_n^{SB}$ can have
small imaginary parts on the imaginary axis, satisfying the left hand
side of \eqref{eq:fourphipole} to a good approximation.  For symmetric
potentials the first radiating component is $\bar\phi_3$.  The
singular breather solution can have an exponentially decaying small
imaginary part on the imaginary axis,
\begin{equation}
\mathrm{Im}\,\bar\phi_3^{SB}=\nu_3\exp(-i\sqrt{8}y)
\ \ \mathrm{for} \quad \mathrm{Re}\,y=0 \,, \label{eq:phi3}
\end{equation}
where $\nu_3$ is some constant. On the other hand, since the
quasibreather solution of the mode equations is regular and symmetric,
$\bar\phi_3^{QB}$ has zero imaginary part on the imaginary axis.

For symmetric potentials the value of $\nu_3$ can be obtained by Borel
summing the series \eqref{eq:yexp1} for $\bar\phi_3$
\cite{Pomeau}. The first step is to define a Borel transformed series
by
\begin{equation}
  V(z)=\sum_{n=2}^{\infty}\frac{\psi_3^{(n)}}{(2n)!}
  z^{2n} \,.  \label{eq:bortr}
\end{equation}
Then the Laplace transform of $V(z)$ will give us the Borel summed
series of $\bar\phi_3^{SB}(y)$ which we denote by
$\hat\phi_3^{SB}(y)$,
\begin{equation}
\hat\phi_3^{SB}(y)=\int_{0}^{\infty} \mathrm{d}t\,
e^{-t}V\left(\frac{t}{y}\right)\,.\label{borelint}
\end{equation}
We are only interested in the imaginary part of $\hat\phi_3^{SB}(y)$
on the negative imaginary axis, $y=-iy_i$, where $y_i>0$ real. Then
the argument of $V$ is $z=t/y=it/y_i$, which is pure imaginary with
positive imaginary part. Since all terms in \eqref{eq:bortr} contain
even powers of $z$, no individual term gives a contribution to ${\rm
  Im}\,\hat\phi_3^{SB}$ on the imaginary axis. The value of ${\rm
  Im}\,\hat\phi_3^{SB}$ is determined there by the leading order large
$n$ behavior of the series \eqref{eq:bortr}.  Using \eqref{eq:p3nser}
and including a term proportional to $z^2$,
\begin{equation}
  V(z)\sim\sum_{n=1}^{\infty}k_D\frac{(-1)^n}{2n}
  \left(\frac{z}{\sqrt 8}\right)^{2n}
=-\frac{k_D}{2}\ln\left(1+\frac{z^2}{8}\right) , \label{e:borelsum}
\end{equation}
where the sign $\sim$ denotes equality up to terms that do not give
contribution to the imaginary part of $\hat\phi_3^{SB}$ on the
imaginary axis. Transforming the argument of the logarithm into
product form, only one of the factors gives a contribution,
\begin{equation}
  V(z)\sim-\frac{k_D}{2}\ln\left(1+\frac{iz}{\sqrt{8}}\right) .
\label{eq:vzln1}
\end{equation}
For purely imaginary $y$,
\begin{equation}
  V\left(\frac{t}{y}\right)\sim
-\frac{k_D}{2}\ln\left(1-\frac{t}{y_i\sqrt{8}}\right) .
\label{eq:vzln2}
\end{equation}
In this case, for $t>y_i\sqrt{8}$ we have to integrate along the
branch cut of the logarithm function. In order to see how to go around
the singularity at $t=y_i\sqrt{8}$ we note that according to
\eqref{eq:matchreg}, the $1/y^2$ expansion
\eqref{eq:yexp1}-\eqref{eq:yexp3} has been applied for $y=y_r-iy_i$,
where $y_r$ and $y_i$ are positive and real. This corresponds to the
requirement of exponential decay for $r>0$ along the real $r$
axis. Then
\begin{equation}
iz=\frac{t}{y_r^2+y_i^2}(-y_i+iy_r) \,,
\end{equation}
which shows that the argument of the logarithm in \eqref{eq:vzln1} has
to go around the singularity in the upper half of the complex plane.
This means that we approach the branch cut of the logarithm at the
negative part of the real axis from above, where its imaginary part is
$\pi$. Then for purely imaginary $y$ we can evaluate the imaginary
part of the integral \eqref{borelint} by integrating on the branch
cut,
\begin{align}
{\rm Im}\,\hat\phi_3^{SB}(y)&=-\int_{i \sqrt8 \,y}^{\infty}
\mathrm{d}t\, e^{-t}\,\frac{k_D\pi}{2} \notag\\
&=-\frac{k_D\pi}{2}\exp\left(-i \sqrt8 \,y\right) .
\label{e:borel}
\end{align}
The logarithmic singularity of $V\left(t/y\right)$ does not contribute
to the integral. Comparing with \eqref{eq:phi3},
\begin{equation}
\nu_3=-\frac12 k_D\pi \,.  \label{eq:nu3kd}
\end{equation}

For asymmetric potentials the leading order radiating component will
be in $\bar\phi_2$, and
\begin{equation}
\mathrm{Im}\,\bar\phi_2^{SB}=\nu_2\exp(-i\sqrt{3}y)
\ \ \mathrm{for} \quad \mathrm{Re}\,y=0 \,.
\end{equation}
Because the dominant behavior of $\bar\phi_0$, it is not possible to
determine the constant $\nu_2$ by the Borel summation. Its value can
be calculated by numerical integration of the Fourier mode equations,
following the method presented in \cite{SK} and \cite{FFHM}.

It is reassuring that even though we work with a truncated set of mode
equations Birkhoff's theorem still holds in the following
sense. Neither $\alpha_{k}^{(n)}$ nor $\beta_{k}^{(n)}$ has an
appropriately singular behavior so that they generate an imaginary
correction on the imaginary axis for $\bar A_{k}$ and $\bar
B_{k}$. The Borel summation procedure does not produce gravitational
radiation.

\subsection{Construction of the quasibreather}

As we have already discussed, both the singular breather (SB), and the
quasibreather (QB) solutions are well approximated in a large domain
by the small-amplitude expansion. Since the tail is exponentially
suppressed in $\varepsilon$, apart from a small central region around
$r=0$, where the SB solution gets too large, the QB and SB solutions
are extremely close to each other. We denote the difference in the
first radiating Fourier component $\bar\phi_3$ of the two solutions by
\begin{equation}
\bar\phi_3^w=\bar\phi_3^{QB}-\bar\phi_3^{SB} \,.
\end{equation}
The small function $\bar\phi_3^w$ solves the linearization of the wave
equation around the singular breather solution. To leading order in
$\varepsilon$ this reduces to the flat background wave equation
\begin{equation}
\frac{d^2\bar\phi_3^w}{dr^2}
+\frac{D-1}{r}\,\frac{d\bar\phi_3^w}{dr}
+8\bar\phi_3^w=0 \,. \label{eq:flatw}
\end{equation}
The general solution of \eqref{eq:flatw} can be written as
\begin{equation}
\bar\phi_3^w=\frac{\sqrt[4]{2}\sqrt{\pi}}{r^{D/2-1}}
\left[\alpha_D^{}Y_{D/2-1}(\sqrt{8}r)
+\beta_D^{}J_{D/2-1}(\sqrt{8}r)
\right] ,  \label{eq:phi3w}
\end{equation}
where $J$ and $Y$ are Bessel functions of the first and second kinds,
and $\alpha_D$, $\beta_D$ are constants. The asymptotic behavior of
the Bessel functions is
\begin{align}
J_\nu(x)&\approx\sqrt{\frac{2}{\pi x}}\cos\left(x-\frac{\nu\pi}{2}
-\frac{\pi}{4}\right) , \\
Y_\nu(x)&\approx\sqrt{\frac{2}{\pi x}}\sin\left(x-\frac{\nu\pi}{2}
-\frac{\pi}{4}\right) , \label{eq:yasympt}
\end{align}
for $\arg x<\pi$ and $|x|\to\infty$. The constants $\alpha_D$ and
$\beta_D$ describe the amplitude of the standing wave tails in
$\bar\phi_3^w$, since for large distances from the center,
\begin{align}
\bar\phi_3^w&\approx\frac{1}{r^{(D-1)/2}}\biggl\{
\alpha_D\sin\left[\sqrt{8}\,r-(D-1)\frac{\pi}{4}\right]
\label{eq:phi3t}\\
&\qquad\qquad\qquad
+\beta_D\cos\left[\sqrt{8}\,r-(D-1)\frac{\pi}{4}\right]
\biggr\} . \notag
\end{align}
Since the SB solution is exponentially decaying, this will also
be the tail of the QB configuration.

The second term in \eqref{eq:phi3w} gives a purely real contribution
to $\bar\phi_3^w$ on the imaginary axis. However, converting
\eqref{eq:yasympt} into exponential form,
\begin{align}
Y_\nu(x)&\approx\frac{1}{\sqrt{2\pi x}}\biggl\{
\exp\left[ix-\frac{i\pi}{4}(2\nu+3)\right] \\
&\qquad\qquad\quad
+\exp\left[-ix+\frac{i\pi}{4}(2\nu+3)\right]
\biggr\} , \notag
\end{align}
we see that the first term in \eqref{eq:phi3w} yields an exponentially
behaving imaginary part along the imaginary axis. Close to the upper
pole at $iQ_D/\varepsilon$, using the coordinate $y$ defined in
\eqref{eq:ry}, to leading order in $\varepsilon$ we obtain that
\begin{equation}
\mathrm{Im}\,\bar\phi_3^w=\frac{\alpha_D}{2}
\left(\frac{\varepsilon}{Q_D}\right)^{\frac{D-1}{2}}
\exp\left(\frac{\sqrt{8}Q_D}{\varepsilon}-i\sqrt{8}\,y\right) ,
\label{eq:phi3wim}
\end{equation}
for $\mathrm{Re}\,y=0$. Since our aim is to obtain a non-singular QB
solution which is symmetric for $r\to-r$ for real $r$,
\eqref{eq:phi3wim} must cancel the exponential behavior of
$\mathrm{Im}\,\bar\phi_3^{SB}$ given by \eqref{eq:phi3}. This fixes
the amplitude $\alpha_D$,
\begin{equation}\label{eq:tampl0}
\alpha_D^{}=-2\nu_3\left(\frac{Q_{D}}{\varepsilon}\right)^{\frac{D-1}{2}}
\exp\left(-\frac{\sqrt{8}Q_{D}}{\varepsilon}\right) \ .
\end{equation}
Substituting the value of $\nu_3$ from \eqref{eq:nu3kd}, obtained by
the Borel summation,
\begin{equation}\label{eq:tampl}
\alpha_D^{}=k_D\pi\left(\frac{Q_{D}}{\varepsilon}\right)^{\frac{D-1}{2}}
\exp\left(-\frac{\sqrt{8}Q_{D}}{\varepsilon}\right) \ .
\end{equation}

For any value of $\beta_D$ the second term in \eqref{eq:phi3w} gives a
regular symmetric contribution to $\bar\phi_3^w$, which does not
change the behavior of the imaginary part on the real axis. However,
as it is apparent from \eqref{eq:phi3t}, any nonzero $\beta_D$
necessarily increases the tail amplitude, and consequently the energy
density in the tail as well. Hence, in order to obtain the minimal
tail quasibreather, we set
\begin{equation}
\beta_D=0 \,.
\end{equation}

The standing wave tail of the quasibreather in the asymptotic region
is given by the first term of \eqref{eq:phi3w},
\begin{align}
\phi^{QB}&=\sqrt[4]{2}\sqrt{\pi}\,
\frac{\alpha_D^{}}{r^{D/2-1}}Y_{D/2-1}
(\sqrt{8}r)\cos(3\tau) \label{eq:qbtail}\\
&\approx \frac{\alpha_D^{}}{r^{(D-1)/2}}
\sin\left[\sqrt{8}r-(D-1)\frac{\pi}{4}\right]\cos(3\tau) \,. \notag
\end{align}
Subtracting the regular solution involving the Bessel function $J_\nu$
with a phase shift in time, we cancel the incoming radiating
component, and obtain the radiative tail of the oscillaton,
\begin{align}
\phi^{osc}&=\sqrt[4]{2}\sqrt{\pi}\,
\frac{\alpha_D^{}}{r^{D/2-1}}\Bigl[Y_{D/2-1}(\sqrt{8}r)\cos(3\tau) \notag\\
&\qquad\qquad\qquad\qquad\
-J_{D/2-1}(\sqrt{8}r)\sin(3\tau)\Bigr]\nonumber\\
&\approx \frac{\alpha_D^{}}{r^{(D-1)/2}}
\sin\left[\sqrt{8}r-(D-1)\frac{\pi}{4}-3\tau \right] . \label{eq:radtail}
\end{align}
Equations \eqref{eq:qbtail} and \eqref{eq:radtail} are valid for
symmetric potentials.  In both cases, the amplitude of the tail of
$\phi$ at large $r$ is given by $\alpha_D^{}$. According to
\eqref{eq:rscpu}, the physical amplitude is
$\alpha_D^{}/\sqrt{8\pi}$. Since the transformation \eqref{eq:trans}
changes the coordinates, $\alpha_D^{}$ scales as $m^{(1-D)/2}$ with
the scalar field mass $m$.

\subsection{Tail amplitude}

The scalar field tail calculated in the previous subsection is so
small that it is not surprising that it has not been detected by
numerically solving the Fourier mode equations in \cite{Seidel1} and
\cite{LopezMatos}. In order to relate the magnitude of the oscillating
tail to the central amplitude, we represent $\phi$ in the core region
by $\phi=\varepsilon^2p_2\cos\tau$, and in the tail by
\eqref{eq:qbtail}. The tail starts to dominate at a radius $r=r_t$
where
\begin{equation}
\phi(\tau=0,r)=\varepsilon^2p_2(\varepsilon
r)=\varepsilon^2S(\varepsilon r)\sqrt{\frac{D-1}{D-2}}
\end{equation}
equals to
$\alpha_D^{}r^{(1-D)/2}$. Since $s\approx -1+s_1\rho^{2-D}$ for large
$\rho$, the asymptotic behavior of $S$ in the relevant dimensions is
\begin{align}
S_{D=3}(\rho)&=S_te^{-\rho}\rho^{s_1/2-1}\left[1-\frac{s_1(s_1-2)}{8\rho}
+\mathcal{O}\left(\frac{1}{\rho^2}\right)\right] , \\
S_{D=4}(\rho)&=S_t\frac{e^{-\rho}}{\rho^{3/2}}\left[1-\frac{4s_1-3}{8\rho}
+\mathcal{O}\left(\frac{1}{\rho^2}\right)\right] , \\
S_{D=5}(\rho)&=S_t\frac{e^{-\rho}}{\rho^{2}}\left[1+\frac{1}{\rho}
-\frac{s_1}{4\rho^2}
+\mathcal{O}\left(\frac{1}{\rho^4}\right)\right] ,
\end{align}
where the constants $s_1$ and $S_t$ are given in Table
\ref{c1table}.
\begin{table}[htbp]
\begin{tabular}{|c|c|c|c|}
\hline
  & $D=3$  & $D=4$  & $D=5$ \\
  \hline
  $s_1$ & 3.505 & 7.695 & 10.40 \\
  $S_t$ & 3.495 & 88.24 & 23.39 \\
  \hline
\end{tabular}
\caption{The numerical values of the constants $s_1$ and $S_t$
  in $3$, $4$ and $5$ spatial dimensions. \label{c1table}}
\end{table}
The values of $r_t$ and the amplitude of the tail at that radius,
$\Phi_t=\alpha_D^{}r^{(1-D)/2}_t/\sqrt{8\pi}$ for several
$\varepsilon$ are given in Table \ref{rttable}.
\begin{table}[htbp]
\begin{tabular}{|c|c|c|c|c|c|c|}
\hline
  & \multicolumn{2}{|c|}{$D=3$}  & \multicolumn{2}{|c|}{$D=4$}
     & \multicolumn{2}{|c|}{$D=5$} \\ \cline{2-7}
  $\varepsilon$ & $r_t$ & $\Phi_t$ & $r_t$ & $\Phi_t$ & $r_t$ & $\Phi_t$\\
  \hline
  $0.1$ & $1160$ & $8.96\cdot10^{-52}$ & $648$ & $2.76\cdot10^{-32}$
    & $346$ & $4.42\cdot10^{-20}$\\
  $0.2$ & $302$ & $4.63\cdot10^{-27}$ & $168$ & $1.06\cdot10^{-17}$
    & $92.6$ & $6.01\cdot10^{-12}$\\
  $0.3$ & $140$ & $9.28\cdot10^{-19}$ & $78.2$ & $9.45\cdot10^{-13}$
    & $45.0$ & $3.83\cdot10^{-9}$\\
  $0.4$ & $81.6$ & $1.40\cdot10^{-14}$ & $46.4$ & $3.07\cdot10^{-10}$
    & $27.9$ & $1.03\cdot10^{-7}$\\
  $0.5$ & $54.3$ & $4.68\cdot10^{-12}$ & $31.5$ & $1.03\cdot10^{-8}$
    & $19.8$ & $7.56\cdot10^{-7}$\\
  $0.6$ & $39.2$ & $2.30\cdot10^{-10}$ & $23.2$ & $1.09\cdot10^{-7}$
    & $15.1$ & $2.87\cdot10^{-6}$\\
  $0.7$ & $29.9$ & $3.76\cdot10^{-9}$ & $18.0$ & $5.91\cdot10^{-7}$
    & $12.2$ & $7.42\cdot10^{-6}$\\
  $0.8$ & $23.8$ & $3.08\cdot10^{-8}$ & $14.6$ & $2.12\cdot10^{-6}$
    & $10.3$ & $1.51\cdot10^{-5}$\\
  \hline
\end{tabular}
\caption{The radius $r_t$ where the oscillating tail starts to
dominate, and its amplitude $\Phi_t$ there. \label{rttable}}
\end{table}

The tail amplitude should be compared to the central amplitude
\begin{equation}
\Phi_c=\varepsilon^2 \Phi_{1c} \ , \qquad
\Phi_{1c}=\frac{S_c}{\sqrt{8\pi}}\sqrt{\frac{D-1}{D-2}} \,,
\end{equation}
where the constants $S_c$ and $\Phi_{1c}$ are given in Table
\ref{sctable}.
\begin{table}[htbp]
\begin{tabular}{|c|c|c|c|}
\hline
  & $D=3$  & $D=4$  & $D=5$ \\
  \hline
  $S_c$ & 1.021 & 3.542 & 14.02 \\
  $\Phi_{1c}$ & 0.288 & 0.865 & 3.229 \\
  $\rho_h$ & 2.218 & 1.357 & 0.763 \\
  \hline
\end{tabular}
\caption{The numerical values of the constants
  $S_c$, $\Phi_{1c}$ and $\rho_h$, which determine the central
  amplitude $\Phi_c$ and the characteristic size $r_h$. \label{sctable}}
\end{table}
The radius $r_t$ where the tail starts to dominate is much larger than
the characteristic radius of the core, which can be defined as the
radius $r_h$ where $\Phi=\Phi_c/2$. Clearly, $r_h=\rho_h/\varepsilon$,
where $\rho_h$ is the value of $\rho$ for which $S=S_c/2$. The value
of $\rho_h$ for various spatial dimensions $D$ is also given in Table
\ref{sctable}. Clearly, there is only some chance to numerically
observe the tail for as large $\varepsilon$ values as $0.5$, which for
$D=3$ is close to the maximum value
$\varepsilon_{\mathrm{max}}\approx0.525$. It can also be observed from
Table \ref{rttable}, that for larger spatial dimensions the radiation
is significantly stronger. It would be reasonable to first make the
numerical analysis for $D=5$, since then the tail has the largest
amplitude. Even if the Klein-Gordon oscillatons are unstable in $D=5$,
as we have seen in Section \ref{sec:totmass}, there are scalar
potentials, for which large amplitude oscillatons are stable. Since
the exponent in \eqref{eq:tampl} is potential independent, in general,
we expect to get a tail amplitude of similar magnitude as for the
Klein-Gordon field.

\subsection{Mass loss rate} \label{sec:massloss}

Since at large distances from the center the mass function $\hat m$
agrees with the total mass $M$, the mass change rate of the oscillaton
can be calculated from the energy current carried by the wave
\eqref{eq:radtail} using \eqref{eq:minkmt}. Averaging for an
oscillation period,
\begin{equation}\label{e:symradlaw}
\frac{\mathrm{d} M}{\mathrm{d} t}=
-\frac{c_1}{m^{D-3}\varepsilon^{D-1}}
\exp\left(-\frac{c_2}{\varepsilon}\right) ,
\end{equation}
where the $D$ dependent constants are
\begin{equation}
c_1=3\sqrt{2} k_D^2Q_{D}^{D-1}
\frac{\pi^{D/2+1}}{4\Gamma\left(\frac{D}{2}\right)}	
 \ , \qquad
c_2=2\sqrt{8}Q_{D} \,.
\end{equation}
The numerical values of $c_1$ and $c_2$ for various spatial dimensions
are listed in Table \ref{ctable}.
\begin{table}[htbp]
\begin{tabular}{|c|c|c|c|}
\hline
  & $D=3$  & $D=4$  & $D=5$ \\
\hline
$c_1$ & $30.0$ & $7.23$ & $0.720$ \\
$c_2$ & $22.4993$ & $13.0372$ & $6.99159$ \\
\hline
\end{tabular}
\caption{\label{ctable}
  The constants  $c_1$ and $c_2$ in the mass
  loss rate expression \eqref{e:symradlaw} for $D=3,4,5$ spatial
  dimensions.
}
\end{table}
The values of $c_2$ are the same for any symmetric potential, but the
numbers given for $c_1$ are valid only for the Klein-Gordon field.

The higher $\varepsilon$ is, the more chance we have to observe the
presumably tiny energy loss.  As we have seen in subsection\
\ref{sec:totmass}, for $D=3$ spatial dimensions oscillons are stable
for $\varepsilon<\varepsilon_{\mathrm{max}}\approx0.525$. The total
mass is maximal at $\varepsilon_{\mathrm{max}}$. Restoring the scalar
field mass $m$ into the expressions, the maximal mass value is
$M_{\mathrm{max}}=0.614/m$. Substituting into \eqref{e:symradlaw}, for
the maximal mass Klein-Gordon oscillaton we get
\begin{equation}
\left(\frac{1}{M}\,\frac{\mathrm{d} M}{\mathrm{d} t}
\right)_{M=M_{\mathrm{max}}}=
-4.3\cdot 10^{-17} m \,.  \label{eq:maxloss1}
\end{equation}
This expression is valid in Planck units. Expressing $M$ in kilograms,
and $mc^2$ in electron volts the maximal oscillaton mass is
\begin{equation}
M_{\mathrm{max}}=0.614\,m_PE_P/m=1.63\cdot 10^{20}kg\frac{eV}{mc^2} \,,
\end{equation}
where the Planck mass is $m_P=2.18\cdot10^{-8}kg$ and the Planck
energy is $E_P=1.22\cdot10^{28}eV$.  Expressing $t$ in seconds, since
$t_p=5.39\cdot 10^{-44}s$, for the maximal mass configuration we get
\begin{equation}
\left(\frac{1}{M}\,\frac{\mathrm{d} M}{\mathrm{d} t}
\right)_{M=M_{\mathrm{max}}}=
-\frac{0.066}{s}\,\frac{mc^2}{eV} \,. \label{eq:maxloss2}
\end{equation}
Note that we only determined the leading order result for the
radiation amplitude. We saw hints that the small-amplitude results
give sensible answers for moderate $\varepsilon$ values, however we do
not have a good control over non-leading terms in the radiation
amplitude. In general for such high $\varepsilon$ values we expect to
have the same exponential factor, but with a different prefactor $c_1$
\cite{FFHM,moredim,dilaton}. The order of magnitude should be
nevertheless correct.

According to \eqref{eq:totmass}, for $D=3$ spatial dimensions, to
leading order the total mass is proportional to the
amplitude. $M=\varepsilon M^{(1)}/m$, where from Table\
\ref{masstable}, $M^{(1)}=1.75266$. Substituting into
\eqref{e:symradlaw}, for the Klein-Gordon case this yields
\begin{equation}
\frac{\mathrm{d} M}{\mathrm{d} t}=
-\frac{c_3}{m^2M^2}\,
\exp\left(-\frac{c_4}{mM}\right) \,. \label{eq:linmassloss}	
\end{equation}
where
\begin{equation}
c_3=92.2 \ , \qquad
c_4=39.4337 \,.	 \label{eq:c3c4}
\end{equation}
Expression \eqref{eq:linmassloss} has the same form as the classical
mass loss formula $(122)$ of \cite{Page}, although the constant
corresponding to $c_3$ is much larger there, it is $3797437.776$. This
means that the amplitude of the radiating tail of the scalar field
$\Phi$ is overestimated by a factor $202.9$ in \cite{Page}. In order
to understand the reason for this large difference, and why it is
necessary to use our more complicated approach to obtain a correct
mass loss rate, we first describe the method of \cite{Page} in our
formalism.

For a $D=3$ Klein-Gordon system let us consider the third Fourier
component \eqref{eq:fourphi} of the wave equation \eqref{eq:wave3} in
the $\omega\to1$ limit. Taking the results
\eqref{eq:bphi1}-\eqref{eq:bb2} of the small-amplitude expansion, and
substituting into the nonlinear terms on right hand side of
\eqref{eq:fourphi}, we obtain an inhomogeneous linear differential
equation for $\bar\phi_3$,
\begin{equation}
\frac{d^2\bar\phi_3}{dr^2}
+\frac{2}{r}\,\frac{d\bar\phi_3}{dr}
+8\bar\phi_3
=P(r) \,.  \label{eq:linphi3}
\end{equation}
Here the function $P(r)$ is given by the small-amplitude expansion, in
a power series form in $\varepsilon$,
\begin{equation}
P(r)=\sum_{k=3}^\infty P_{2k}(r)\varepsilon^{2k} \,.
\end{equation}
For the $D=3$ Klein-Gordon system the leading order term is
\begin{equation}
P_6(r)=\frac{3}{16}p_2^3(\varepsilon r)
+p_2a_4^{(2)}(\varepsilon r) \,.  \label{eq:pp6d3}
\end{equation}
Equation \eqref{eq:linphi3} with $P(r)=\varepsilon^6P_6(r)$, but
setting $a_4^{(2)}=0$, corresponds to equation $(52)$ of
\cite{Page}. Since there the Fourier modes are defined in terms of
exponentials instead of cosine functions, the coefficient of the
$p_2^3$ term is $3/4$ in \cite{Page} instead of $3/16$. The term
containing $a_4^{(2)}$ is missing there because of the assumption that
the $g_{tt}=-A$ metric component is time independent. However, at
$\varepsilon^4$ order either $g_{tt}$ becomes oscillatory, or the
spatial metric ceases to be conformally flat.

The oscillating tail responsible for the radiation loss in the
$\bar\phi_3$ mode can be estimated by integrating \eqref{eq:linphi3}
using the Green function method,
\begin{align}
\bar\phi_3(r)&=\frac{\cos(\sqrt{8}r)}{\sqrt{8}r}
\int_0^r\bar r\sin(\sqrt{8}\bar r)P(\bar r)d\bar r \\
&\quad+\frac{\sin(\sqrt{8}r)}{\sqrt{8}r}
\int_r^\infty\bar r\cos(\sqrt{8}\bar r)P(\bar r)d\bar r \,. \notag
\end{align}
The oscillaton core is exponentially localized, hence in the tail
region it is a very good approximation to write
\begin{equation}
\bar\phi_3(r)=\bar\alpha\frac{\cos(\sqrt{8}r)}{r}  \,,
\end{equation}
where the constant determining the amplitude is
\begin{equation}
\bar\alpha=\frac{1}{\sqrt{8}}
\int_0^\infty r\sin(\sqrt{8} r)P(r)dr  \,.
\end{equation}
Since the functions describing the oscillaton by the $\varepsilon$
expansion are symmetric around $r=0$, and since the sine function can
be written as the difference of two exponentials,
\begin{equation}
\bar\alpha=\frac{1}{2i\sqrt{8}}
\int_{-\infty}^\infty r\exp(\sqrt{8}ir)P(r)dr  \,.
\end{equation}
The functions in the small-amplitude expansion depend directly on the
rescaled radial coordinate $\rho=\varepsilon r$, so it is natural to
write the integral into the form
\begin{equation}
\bar\alpha=\frac{1}{2i\sqrt{8}\varepsilon^2}
\int_{-\infty}^\infty\rho
\exp\left(\frac{\sqrt{8}i\rho}{\varepsilon}\right)
P\left(\frac{\rho}{\varepsilon}\right)d\rho  \,. \label{eq:alphaintrho}
\end{equation}
This can be replaced by a contour integral around the upper plane, and
can be approximated by taking into account the pole which is closest
to the real axis.  The position of the closest pole is the same as that
of the SN equations \eqref{Seq} and \eqref{seq}, it is at $\rho=iQ_3$
on the imaginary axis. Let us first calculate the contribution from
the leading $P_6$ term, given by \eqref{eq:pp6d3}. Then, since for
three spatial dimensions $p_2=\sqrt{2}S$, using \eqref{eq:sexp} and
\eqref{eq:a42r}, the behavior near the pole is
\begin{equation}
P\left(\frac{\rho}{\varepsilon}\right)\approx\varepsilon^6
P_6\left(\frac{\rho}{\varepsilon}\right)\approx
-\frac{3^5\sqrt{2}\varepsilon^6}{5R^6} \,, \label{eq:p6pole}
\end{equation}
where $R=iQ_3-\rho$. The omission of the term containing $a_4^{(2)}$
from \eqref{eq:pp6d3} results in a value which is $5/3$ times that of
\eqref{eq:p6pole}. The residue can be calculated by integrating by
parts five times,
\begin{equation}
\bar\alpha=-\frac{2^33^4\sqrt{2}\pi Q_3}{5^2\varepsilon}
\exp\left(-\frac{\sqrt{8}Q_3}{\varepsilon}\right)  \,.
\end{equation}
This amplitude is much larger than the amplitude $\alpha_3$ calculated
by the Borel summation method in \eqref{eq:tampl},
\begin{equation}
\frac{\bar\alpha}{\alpha_3}=-\frac{2^43^4\sqrt{2}}{5^2k_3}
\approx 121.8 \,.
\end{equation}
If we omit the term containing $a_4^{(2)}$ from \eqref{eq:pp6d3}, we
have a $5/3$ factor, and get $\bar\alpha/\alpha_3=202.9$, which is the
ratio of the tail amplitude of \cite{Page} to our value, as we have
mentioned after Eq.\ \eqref{eq:c3c4}.

The fundamental problem with the above calculated tail amplitude
$\bar\alpha$ is that it is just the first term of an infinite series,
of which all terms give contributions which are the same order in
$\varepsilon$. This can be illustrated by calculating the contribution
of the next term in $P(r)$.  Although the small-amplitude expansion
yields a rather complicated expression for $P_8(r)$, it contains terms
proportional to $p_2^3a_2$, $p_4a_4^{(2)}$ and $p_2a_2a_4^{(2)}$,
which have eighth order poles. When calculating the integral
\eqref{eq:alphaintrho} one has to integrate by parts seven times, so
the result will have the same $\varepsilon$ order as the earlier
result calculated from the $P_6(r)$ term. Even if we could calculate
higher order contributions, we have no reason to expect that the
series converges, and even if it would be convergent it may not give a
correct result for the mass loss. This has already been demonstrated
for the simpler system of a real scalar field with a nontrivial
interaction potential on flat Minkowski background. It was first
pointed out in \cite{Eleonski} that there are too many boundary
conditions to satisfy when solving the Fourier mode equations in order
to find periodic localized breather solutions. In \cite{Eleonski} the
energy loss rate of the long living oscillon configurations was
estimated by a method analogous to that of \cite{Page}. However,
calculating higher order contributions, it turned out that the method
gives an incorrect nonzero result even for the periodic sine-Gordon
breather. Moreover, the expansion is not convergent for the $\phi^4$
scalar theory. The proper approach to calculate the lifetime of
oscillons has been worked out by \cite{SK} and \cite{Pomeau}, using
complex extension and Borel summation. As a result of the above
arguments, the correct value of $c_3$ in the mass loss rate expression
\eqref{eq:linmassloss} for the $D=3$ Klein-Gordon field is $c_3=92.2$.

Although the expression \eqref{eq:linmassloss} is correct for small
values of $M$, since it is based on the assumption that $M$ depends
linearly on $\varepsilon$, one should not apply it to mass values
close to $M_{\mathrm{max}}$. For example, substituting the value of
the maximal mass $M_{\mathrm{max}}=0.614/m$ into
\eqref{eq:linmassloss}, we obtain
\begin{equation}
\left(\frac{1}{M}\,\frac{\mathrm{d} M}{\mathrm{d} t}
\right)_{M=M_{\mathrm{max}}}=
-5.0\cdot 10^{-26} m \,,  \label{eq:linmaxloss}
\end{equation}
which is $9$ magnitudes smaller than the maximal mass loss rate
obtained in \eqref{eq:maxloss1}. The reason for this huge difference
is that according to the linear expression $M=\varepsilon M^{(1)}$, to
the mass value $M_{\mathrm{max}}=0.614/m$ belongs an $\varepsilon$
value of $0.350$. At that $\varepsilon$ we obviously get a
significantly lower radiation than at
$\varepsilon_{\mathrm{max}}\approx0.525$, because of the exponential
dependence. Since expression \eqref{eq:maxloss1} does not involve this
approximation, we expect it to give a more reliable result.


\subsection{Time dependence}

Instead of using \eqref{eq:linmassloss} to determine the time
dependence of the oscillaton mass, in order to obtain results that are
valid for larger mass values, we work out a method involving a higher
order approximation for the $\varepsilon$ dependence of the mass.
Since the first two terms of \eqref{eq:totmass} determine the mass
maximum to a good precision, we expect it to be a reasonable
approximation for close to maximal $\varepsilon$ values. Including the
scalar field mass $m$, we use
\begin{equation}
M=\varepsilon^{4-D}m^{2-D}
\left(M^{(1)}+\varepsilon^2M^{(2)}\right) \,. \label{eq:massord2}
\end{equation}
Taking the time derivative and comparing with \eqref{e:symradlaw},
\begin{equation}
\frac{dt}{d\varepsilon}=-\frac{\varepsilon^2}{m}
\left(\beta_1+\beta_2\varepsilon^2\right)
\exp\left(\frac{c_2}{\varepsilon}\right) ,
\end{equation}
where
\begin{equation}
\beta_1=\frac{4-D}{c_1}M^{(1)} \ , \qquad
\beta_2=\frac{6-D}{c_1}M^{(2)} \,.
\end{equation}
This can be integrated in terms of the exponential integral function,
\begin{align}
t-t_0&=-\frac{\varepsilon}{120m}\bigl[
20\beta_1\left(c_2^2+c_2\varepsilon+2\varepsilon^2\right) \notag\\
&+\beta_2\left(c_2^4+c_2^3\varepsilon+2c_2^2\varepsilon^2
+6c_2\varepsilon^3+24\varepsilon^4\right)
\bigr]\exp\left(\frac{c_2}{\varepsilon}\right) \notag\\
&+\frac{c_2^3}{120m}\left(20\beta_1+\beta_2c_2^2\right)
\mathrm{Ei}\left(\frac{c_2}{\varepsilon}\right) .
\end{align}
Taking the expansion of the result, for small $\varepsilon$,
\begin{align}
t-t_0&=
\frac{\varepsilon^4}{m}\biggl[\frac{\beta_1}{c_2}
+\frac{4\beta_1}{c_2^2}\varepsilon  \label{eq:tt0s}\\
&+\frac{20\beta_1+\beta_2c_2^2}{c_2^3}
\left(\varepsilon^2+\frac{6\varepsilon^3}{c_2}
+\mathcal{O}\left(\varepsilon^4\right)\right)
\biggr]\exp\left(\frac{c_2}{\varepsilon}\right) . \notag
\end{align}
Although the correction from the subleading term $M^{(2)}$ only
appears in the third term in the bracket, its influence for
$\varepsilon\approx0.5$ is not negligible. It can be also seen that
for $D=4$ we have $\beta_1=0$, and \eqref{eq:tt0s} starts with an
$\varepsilon^6$ term. The elapsed time as a function of the oscillaton
mass can be obtained by expressing $\varepsilon$ from
\eqref{eq:massord2} and substituting into \eqref{eq:tt0s}.

For $D=3$ spatial dimensions it is natural to start with a maximal
mass configuration $M=M_{\mathrm{max}}$, and wait for the mass to
decrease until the ratio $M/M_{\mathrm{max}}$ reaches a given
value. Since the elapsed time $t$ is inversely proportional to the
scalar field mass $m$, in table \ref{ttable} we list the product $tm$.
\begin{table}[htbp]
\begin{tabular}{|l|c|c|c|}
\hline
$\frac{M_{\mathrm{max}}-M}{M_{\mathrm{max}}}$ & $\varepsilon$
  & $tm$
  & $\frac{t}{\mathrm{year}}\,\frac{mc^2}{eV}$ \\
\hline
$0.01$ & $0.482$ & $5.35\cdot 10^{16}$ & $1.12\cdot 10^{-6}$ \\
$0.1$ & $0.383$ & $3.00\cdot 10^{21}$ & $6.26\cdot 10^{-2}$ \\
$0.2$ & $0.320$ & $1.50\cdot 10^{26}$ & $3.12\cdot 10^{3}$ \\
$0.3$ & $0.269$ & $4.42\cdot 10^{31}$ & $9.22\cdot 10^{8}$ \\
$0.31884$ & $0.260$ & $6.57\cdot 10^{32}$ & $1.37\cdot 10^{10}$ \\
$0.4$ & $0.224$ & $3.99\cdot 10^{38}$ & $8.32\cdot 10^{15}$ \\
$0.5$ & $0.182$ & $1.22\cdot 10^{48}$ & $2.55\cdot 10^{25}$ \\
$0.6$ & $0.144$ & $1.28\cdot 10^{62}$ & $2.67\cdot 10^{39}$ \\
$0.7$ & $0.107$ & $1.94\cdot 10^{85}$ & $4.04\cdot 10^{62}$ \\
\hline
\end{tabular}
\caption{\label{ttable}
The time necessary for the oscillaton mass to decrease to $M$ from the
value $M_{\mathrm{max}}$ at $t=0$. The value of $tm$ is given in
Planck units, and also when the time is
measured in years and the scalar mass in electron volts.
}
\end{table}

Next we address the question that how much of its mass an initially
maximal mass oscillaton loses during the age of the universe, which we
take to be $1.37\cdot 10^{10}$ years. In Table \ref{mstable} we list
the resulting oscillaton masses in units of solar masses $(M_\odot)$,
as a function of the scalar field mass in $eV/c^2$ units.
\begin{table}[htbp]
\begin{tabular}{|c|c|c|c|}
\hline
$\frac{mc^2}{eV}$ & $\varepsilon_{\mathrm{max}}-\varepsilon$
  & $\frac{M}{M_\odot}$
  & $\frac{M_{\mathrm{max}}-M}{M_{\mathrm{max}}}$ \\
\hline
$10^{-35}$ & $5.09\cdot 10^{-20}$
   & $8.20\cdot 10^{24}$ & $1.41\cdot 10^{-38}$ \\
$10^{-30}$ & $5.09\cdot 10^{-15}$
   & $8.20\cdot 10^{19}$ & $1.41\cdot 10^{-28}$ \\
$10^{-25}$ & $5.09\cdot 10^{-10}$
   & $8.20\cdot 10^{14}$ & $1.41\cdot 10^{-18}$ \\
$10^{-20}$ & $5.08\cdot 10^{-5}$
   & $8.20\cdot 10^{9}$ & $1.40\cdot 10^{-8}$ \\
$10^{-15}$ & $0.0704$
   & $7.99\cdot 10^{4}$ & $0.0258$ \\
$10^{-10}$ & $0.163$
   & $7.14\cdot 10^{-1}$ & $0.129$ \\
$10^{-5}$ & $0.223$
   & $6.30\cdot 10^{-6}$ & $0.232$ \\
$1$ & $0.266$
   & $5.58\cdot 10^{-11}$ & $0.319$ \\
$10^{5}$ & $0.297$
   & $5.00\cdot 10^{-16}$ & $0.390$ \\
$10^{10}$ & $0.322$
   & $4.52\cdot 10^{-21}$ & $0.449$ \\
$10^{15}$ & $0.342$
   & $4.12\cdot 10^{-26}$ & $0.498$ \\
\hline
\end{tabular}
\caption{\label{mstable}
  Mass $M$ of an initially maximal mass oscillaton after a period
  corresponding to the age of the universe for various scalar
  field masses. The decrease in $\varepsilon$ from
  $\varepsilon_{\mathrm{max}}=0.525$, and the relative mass change rate
  $(M_{\mathrm{max}}-M)/M_{\mathrm{max}}$ is also given.
}
\end{table}
In order to facilitate comparison, we have chosen the same scalar
field masses as in Eq.\ (178) of \cite{Page}. The first two orders of
the small-amplitude expansion yielded $mM_{\mathrm{max}}=0.614$ in
Planck units for the maximal mass of the oscillaton. Taking the scalar
mass in electron volts, this corresponds to
$M_{\mathrm{max}}=8.20\cdot 10^{-11}M_\odot eV/(mc^2)$, which was used
in Table \ref{mstable}. The value $mM_{\mathrm{max}}=0.607$ from the
numerical solution of the Fourier mode equations calculated in
\cite{Lopez} corresponds to $M_{\mathrm{max}}=8.11\cdot
10^{-11}M_\odot eV/(mc^2)$ in natural units, which is the value used
in \cite{Page}.  Comparing our Table \ref{mstable} to the numbers in
(178) of \cite{Page}, after compensating for the shift in the initial
mass, it is apparent, that for small scalar field masses, i.e.\ for
$m\leq10^{-10}eV/c^2$, oscillatons decay more slowly in
\cite{Page}. The reason for this is that \cite{Page} uses a linear
dependence of the mass on the small parameter, and consequently
underestimates the radiation rate close to the maximum mass, similarly
as we did in \eqref{eq:linmaxloss}. For $m\geq10^{-5}eV/c^2$
oscillatons radiate faster in \cite{Page}, which is a consequence of
the much larger value of the constant $c_3$ in the mass loss law
\eqref{eq:linmassloss} used there. In spite of the differences, the
overall picture remains essentially the same. For all scalar field
masses that appear physically reasonable, a maximal mass oscillaton
loses a significant part of its mass during the lifetime of the
universe. This mass decrease is greater than $10\%$ if $m>4.57\cdot
10^{-12}eV/c^2$, but it remains below $50\%$ if $m<1.85\cdot
10^{15}eV/c^2$.  The above results support the possibility that
provided a scalar field exist in Nature, at least some of the dark
matter content of our Universe would be in the form of oscillatons.

\section{Conclusions}

We have derived an infinite set of radial ODEs determining the spatial
field profiles of bounded solutions of time-dependent, spherically
symmetric Einstein-scalar field equations in the limit when the scalar
field amplitude tends to zero.  The lowest order equations are nothing
but the D-dimensional generalization of the Schr\"odinger-Newton (SN)
eqs. The SN eqs.\ admit globally regular, exponentially decreasing
solutions for spatial dimensions $2<D<6$. The eqs.\ corresponding to
higher orders in the expansion are linear inhomogenous ODEs.  The
class of solutions we are interested in are oscillatons, which loose
slowly their mass by scalar radiation.  In the small-amplitude
expansion we have obtained an asymptotic series for the spatially well
localized core of oscillatons and related their radiation amplitude to
that of the standing wave tail of exactly time-periodic
quasibreathers.  For the class of symmetric scalar potentials we have
determined the amplitude of the standing wave tail of time-periodic
quasibreathers analytically adapting the method of Segur-Kruskal and
using Borel summation.  We have explicitly computed the mass loss rate
for the Einstein-Klein-Gordon system in $D=3,4,5$.

\appendix

\begin{widetext}
\section{Einstein tensor and the wave equation}\label{app:einstein}

The components of the Einstein tensor in the general spherically
symmetric coordinate system \eqref{eq:metrgen} are
\begin{align}
G_{tt}&=(D-1)\left\{\frac{B_{,t}C_{,t}}{4BC}
-\frac{A}{4C_{,r}}\left[\frac{\left(C_{,r}\right)^2}{BC}\right]_{,r}
+(D-2)\frac{A}{2C}\left[1+\frac{\left(C_{,t}\right)^2}{4AC}
-\frac{\left(C_{,r}\right)^2}{4BC}\right]
\right\}
, \label{eq:gtt}\\
G_{rr}&=(D-1)\left\{\frac{A_{,r}C_{,r}}{4AC}
-\frac{B}{4C_{,t}}\left[\frac{\left(C_{,t}\right)^2}{AC}\right]_{,t}
-(D-2)\frac{B}{2C}\left[1+\frac{\left(C_{,t}\right)^2}{4AC}
-\frac{\left(C_{,r}\right)^2}{4BC}\right]
\right\}
, \\
G_{tr}&=
-(D-1)\left[
\frac{A}{4\sqrt{C}}\left(\frac{C_{,t}}{A\sqrt{C}}\right)_{,r}
+\frac{B}{4\sqrt{C}}\left(\frac{C_{,r}}{B\sqrt{C}}\right)_{,t}
\right] , \\
G_{\theta_1\theta_1}&=\frac{C}{4A_{,r}}
\left[\frac{\left(A_{,r}\right)^2}{AB}\right]_{,r}
-\frac{C}{4B_{,t}}
\left[\frac{\left(B_{,t}\right)^2}{AB}\right]_{,t}
-(D-2)\Biggl\{
1+\frac{1}{4BC_{,t}}\left[\frac{B\left(C_{,t}\right)^2}{A}\right]_{,t}
\notag\\
&-\frac{1}{4AC_{,r}}\left[\frac{A\left(C_{,r}\right)^2}{B}\right]_{,r}
+\frac{1}{2}(D-5)\left[1+\frac{\left(C_{,t}\right)^2}{4AC}
-\frac{\left(C_{,r}\right)^2}{4BC}\right]
\Biggr\}, \label{eq:gthth}\\
G_{\theta_n\theta_n}&=G_{\theta_1\theta_1}\prod_{k=1}^{n-1}\sin^2\theta_k \ .
\end{align}
The wave equation \eqref{eq:wave} takes the form
\begin{equation}
\frac{\phi_{,rr}}{B}-\frac{\phi_{,tt}}{A}
+\frac{\phi_{,r}}{2AC^{D-1}}\left(\frac{AC^{D-1}}{B}\right)_{,r}\\
-\frac{\phi_{,t}}{2BC^{D-1}}\left(\frac{BC^{D-1}}{A}\right)_{,t}
-\bar U'(\phi)=0 \,. \label{eq:wave2}
\end{equation}
\end{widetext}

\section{Small-amplitude expansion in Schwarzschild coordinates}
\label{app:schw}

In the main part of the paper we have used the spatially conformally
flat coordinate system $C=r^2B$. In this appendix we present the
results of the $\varepsilon$ expansion in $C=r^2$ Schwarzschild area
coordinates, in order to compare and to point out the disadvantages.
The time dependence of the scalar field $\phi$ and the metric
components $A$ and $B$ up to $\varepsilon^2$ order are
\begin{align}
\phi&=\varepsilon^2p_2\cos\tau
+\mathcal{O}(\varepsilon^4) \,, \\
A&=1+\varepsilon^2a_2+\varepsilon^2a_2^{(2)}\cos(2\tau)
+\mathcal{O}(\varepsilon^4)
\,, \label{eq:a2sch}\\
B&=1+\varepsilon^2b_2
+\mathcal{O}(\varepsilon^4)
\,,
\end{align}
where $p_2$, $a_2$, $a_2^{(2)}$ and $b_2$ are functions of $\rho$.
The functions $a_2$ and $p_2$ are again determined by the coupled
differential equations \eqref{eq:x2} and \eqref{eq:p2}, resulting in
the Schr\"odinger-Newton equations. However, $b_2$ is determined as
\begin{equation}
b_2=\frac{\rho}{D-2}\,\frac{da_2}{d\rho} \,,
\end{equation}
instead of \eqref{eq:b2}. The most important difference is the
appearance of the $\cos(2\tau)$ term in \eqref{eq:a2sch}, causing an
$\varepsilon^2$ order oscillation in the metric component $g_{tt}$. In
spatially conformally flat coordinates there are only $\varepsilon^4$
order oscillating terms in the metric components. The amplitude of the
oscillation is determined by the field equations as
\begin{equation}
a_2^{(2)}=-a_2-b_2 \,.
\end{equation}
Substituting into the expression \eqref{eq:acc} of the magnitude of
the acceleration of constant $(r,\theta_1,\theta_2 ...)$ observers, to
leading order we get
\begin{equation}
\mathrm{a}=\frac{\varepsilon^3}{2}\left(\frac{da_2}{d\rho}
+\frac{da_2^{(2)}}{d\rho}\cos(2\tau)\right) .
\end{equation}

\begin{acknowledgments}

This research has been supported by OTKA Grants No. K61636,
NI68228, and by the U.S. Department of Energy (D.O.E.) under
cooperative research agreement DE-FG 0205ER41360.
\end{acknowledgments}


\begin{thebibliography}{99}


\bibitem{Seidel1} E. Seidel and W-M. Suen, \emph{Phys. Rev. Lett.}
  \textbf{66}, 1659 (1991).

\bibitem{Seidel2} E. Seidel and W-M. Suen, \emph{Phys. Rev. Lett.}
  \textbf{72}, 2516 (1994).

\bibitem{Dashen} R. F. Dashen, B. Hasslacher and A. Neveu,
  \emph{Phys. Rev. D} \textbf{11}, 3424 (1975).

\bibitem{BogMak2} I.~L.~Bogolyubskii, and V.~G.~Makhan'kov,
{\em JETP~Letters} {\bf 25}, 107 (1977).

\bibitem{CopelGM95} E.~J.~Copeland, M.~Gleiser and H.-R.~M\"uller,
 {\em Phys.~Rev. D} {\bf 52}, 1920 (1995).

\bibitem{Chris} P.~L.~Christiansen, N.~Gronbech-Jensen, P.~S.~Lomdahl
  and B.~A.~Malomed, {\em Physica Scripta} {\bf 55}, 131 (1997).

\bibitem{PietteZakr98} B.~Piette, W.~J.~Zakrzewski,
{\em Nonlinearity} {\bf 11}, 1103 (1998).

\bibitem{Honda} E.~P.~Honda and M.~W.~Choptuik,
{\em Phys.~Rev. D} {\bf 65}, 084037 (2002).

\bibitem{Hindmarsh-Salmi06} M.~Hindmarsh and P.~Salmi,
{\em Phys.~Rev. D} {\bf 74}, 105005 (2006).

\bibitem{SafTra} P.~M.~Saffin and A.~Tranberg, {\em JHEP} 01(2007)030 (2007).

\bibitem{fggirs} E.~Farhi, N.~Graham, A.~H.~Guth, N.~Iqbal,
  R.~R.~Rosales and N.~Stamatopoulos,
{\em Phys.~Rev. D} {\bf 77}, 085019 (2008).

\bibitem{sicilia2}
M. Gleiser and D. Sicilia,
{\em Phys.~Rev. D.} {\bf 80}, 125037 (2008)

\bibitem{FFGR} G. Fodor, P. Forg\'acs, P. Grandcl\'ement and
  I. R\'acz, \emph{Phys. Rev. D} \textbf{74}, 124003 (2006).

\bibitem{FFHL} G. Fodor, P. Forg\'acs, Z. Horv\'ath and \'A.
  Luk\'acs, \emph{Phys. Rev. D} \textbf{78}, 025003 (2008).

\bibitem{FFHM} G. Fodor, P. Forg\'acs, Z. Horv\'ath and M. Mezei, {\em
    Phys. Rev. D} \textbf{79}, 065002 (2009).

\bibitem{moredim} G. Fodor, P. Forg\'acs, Z. Horv\'ath and M. Mezei,
  \emph{Phys. Lett. B} \textbf{674}, 319-324 (2009).

\bibitem{Kolb:1993hw}
E.~W. Kolb and I.~I. Tkachev,
  {\em Phys. Rev. D} {\bf 49},  5040 (1994).

\bibitem{Khlopov} I.~Dymnikova, L.~Koziel, M.~Khlopov, S. Rubin,
{\em Gravitation and Cosmology} {\bf 6}, 311 (2000).

\bibitem{Broadhead:2005hn}
M. Broadhead and J. McDonald, {\em Phys. Rev. D} \textbf{72}, 
043519 (2005).

\bibitem{Gleiser:2007ts}
M. Gleiser, B. Rogers and J. Thorarinson,  {\em Phys. Rev. D} {\bf 77},
023513 (2008).

\bibitem{Borsanyi2}
Sz. Borsanyi, M. Hindmarsh,
{\em Phys. Rev. D} {\bf 79}, 065010 (2009).

\bibitem{Farhi05} E.~Farhi, N.~Graham, V.~Khemani, R.~Markov and
 R.~Rosales, {\em Phys.~Rev. D} {\bf 72}, 101701(R) (2005).

\bibitem{Graham07a} N.~Graham
{\em Phys.~Rev.~Lett.} {\bf 98}, 101801 (2007).

\bibitem{Graham07b} N.~Graham
{\em Phys.~Rev. D} {\bf 76}, 085017 (2007).

\bibitem{dilaton} G. Fodor, P. Forg\'acs, Z. Horv\'ath and M. Mezei,
  JHEP08(2009)106, (2009).

\bibitem{Ruffini} R. Ruffini and S. Bonazzola, {\em Phys. Rev.}
  \textbf{187}, 1767 (1969).

\bibitem{Friedberg} R. Friedberg, T. D. Lee, and Y. Pang, 
{\em Phys. Rev. D} \textbf{35}, 3640 (1987).

\bibitem{Ferrell} R. Ferrell and M. Gleiser, {\em Phys. Rev. D} 
\textbf{40}, 2524 (1989).

\bibitem{Moroz} I. M. Moroz, R. Penrose and P. Tod,
  \emph{Class. Quantum Grav.} \textbf{15}, 2733 (1998).

\bibitem{Tod} P. Tod and I. M. Moroz, \emph{Nonlinearity} \textbf{12},
  201 (1999).

\bibitem{Page} D. N. Page, \emph{Phys. Rev. D} \textbf{70}, 023002
  (2004).

\bibitem{SK} H. Segur and M. D. Kruskal, \emph{Phys. Rev. Lett.}
  \textbf{58}, 747 (1987).

\bibitem{Feinblum}D. A. Feinblum and W. A. McKinley, {\em Phys. Rev.}
  \textbf{168}, 1445 (1968).

\bibitem{Kaup} D. J. Kaup, {\em Phys. Rev.}
  \textbf{172}, 1331 (1968).

\bibitem{Harrison} B. K. Harrison, K. S. Thorne, M. Wakano and
  J. A. Wheeler, \emph{Gravitation Theory and Gravitational Collapse,}
  University of Chicago Press (1965).

\bibitem{Jetzer} P. Jetzer, \emph{Phys. Rep.}
  \textbf{220}, 163 (1992).

\bibitem{SchunkMielke} F. E. Schunk and E. W. Mielke,
  \emph{Class. Quantum Grav.}  \textbf{20}, R301 (2003).

\bibitem{Hawley} S. H. Hawley and M. W. Choptuik, \emph{Phys. Rev. D}
  \textbf{67}, 024010 (2003).

\bibitem{Hawleyphd} S. H. Hawley, \emph{Scalar analogues of compact
    astrophysical systems, Ph.D. Dissertation,} University of Texas at
  Austin (2000).

\bibitem{Iwazaki1} A. Iwazaki, \emph{Phys. Lett. B}
  \textbf{451}, 123 (1999).

\bibitem{Iwazaki2} A. Iwazaki, \emph{Phys. Rev. D}
  \textbf{60}, 025001 (1999).

\bibitem{Iwazaki3} A. Iwazaki, \emph{Phys. Lett. B}
  \textbf{455}, 192 (1999).

\bibitem{MatosGuzman} T. Matos and F. S. Guzm\'an,
  \emph{Class. Quantum Grav.}  \textbf{18}. 5055 (2001).

\bibitem{Alcubgalactic} M. Alcubierre, F. S. Guzm\'an, T. Matos,
  D. N\'u\~nez and L. A. Ure\~na-L\'opez and P. Wiederhold,
  \emph{Class. Quantum Grav.}  \textbf{19}. 5017 (2002).

\bibitem{Susperregi} M. Susperregi, \emph{Phys. Rev. D} \textbf{68},
  123509 (2003).

\bibitem{Guzman1} F. S. Guzm\'an and L. A. Ure\~na-L\'opez, {\em
    Phys. Rev. D} \textbf{68}, 024023 (2003).

\bibitem{Guzman2} F. S. Guzm\'an and L. A. Ure\~na-L\'opez, {\em
    Phys. Rev. D} \textbf{69}, 124033 (2004).

\bibitem{Hernandez} X. Hern\'andez, T. Matos, R. A. Sussman and
  Y. Verbin, {\em Phys. Rev. D} \textbf{70}, 043537 (2004).

\bibitem{Guzman3} F. S. Guzm\'an and L. A. Ure\~na-L\'opez, {\em
    Astrophys. J.} \textbf{645}, 814 (2006).

\bibitem{Bernal} A. Bernal and F. S. Guzm\'an, {\em Phys. Rev. D}
  \textbf{74}, 063504 (2006).

\bibitem{Lopez} L. A. Ure\~na-L\'opez, \emph{Class. Quantum Grav.}
  \textbf{19}, 2617 (2002).

\bibitem{LopezMatos} L. A. Ure\~na-L\'opez, T. Matos and R. Becerril,
  \emph{Class. Quantum Grav.} \textbf{19}, 6259 (2002).

\bibitem{Kichena2} S. Kichenassamy, \emph{Class. Quantum Grav.}
  \textbf{25}. 245004 (2008).

\bibitem{Diosi} L. Di\'osi, \emph{Phys. Lett. A} \textbf{105}. 199
  (1984).

\bibitem{Penrose} R. Penrose, \emph{Phil. Trans. R. Soc.}
  \textbf{356}. 1927 (1998).

\bibitem{Alcub} M. Alcubierre, R. Becerril, F. S. Guzm\'an, T. Matos,
D. N\'u\~nez and L. A. Ure\~na-L\'opez, \emph{Class. Quantum Grav.}
  \textbf{20}. 2883 (2003).

\bibitem{Brady} P. R. Brady, C. M. Chambers and S. M. C. V. Gon\c
  calves, \emph{Phys. Rev. D} \textbf{56}, R6057 (1997).

\bibitem{Garfinkle} D. Garfinkle, R. Mann and C. Vuille,
  \emph{Phys. Rev. D} \textbf{68}, 064015 (2003).

\bibitem{Balak}
J. Balakrishna, R. Bondarescu, G. Daues and M. Bondarescu,
\emph{Phys. Rev. D} \textbf{77}, 024028 (2008).

\bibitem{Obregon} O. Obreg\'on,  L. A. Ure\~na-L\'opez and
  F. E. Schunck, \emph{Phys. Rev. D} \textbf{72}, 024004 (2005).

\bibitem{Becerril} R. Becerril, T. Matos and L. A. Ure\~na-L\'opez,
  \emph{Gen. Relativ. Gravit.}  \textbf{38}. 633 (2006).

\bibitem{Kodama} H. Kodama, \emph{Prog. Theor. Phys} \textbf{63}, 1217
  (1980).

\bibitem{Hayward} S. A. Hayward, \emph{Phys. Rev. D} \textbf{53}, 1938
  (1996).

\bibitem{MisnerSharp} C. W. Misner and D. H. Sharp, \emph{Phys. Rev.}
  \textbf{136}, B571 (1964).

\bibitem{Kichenassamy} S. Kichenassamy, \emph{Comm. Pur. Appl. Math.}
  \textbf{44}, 789 (1991).

\bibitem{Choquard} P. Choquard, J. Stubbe and M. Vuffray,
  \emph{Differential and Integral Equations}, \textbf{21} 665 (2008).

\bibitem{Tangherlini} F. R. Tangherlini, \emph{Il Nuovo Cimento}
  \textbf{27}, 636 (1963).

\bibitem{leepang} T. D. Lee and Y. Pang, \emph{Nucl. Phys.}
  \textbf{B315}, 477 (1989).

\bibitem{Pomeau} Y. Pomeau, A. Ramani and B. Grammaticos,
  \emph{Physica} \textbf{D31}, 127 (1988).

\bibitem{Eleonski} V. M. Eleonski, N. E. Kulagin, N. S. Novozhilova
  and V. P. Silin, \emph{Teor. Mat. Fiz} \textbf{60}, 896 (1984).

\end{thebibliography}
\end{document}